\documentclass[12pt]{article}

\title{Set-valued risk measures for conical market models}

\author{Andreas H. Hamel\thanks{Yeshiva University, Department of Mathematical Sciences, New York, NY 10033, USA,
hamel@yu.edu}, Frank Heyde\thanks{University Halle-Wittenberg, Institute of Mathematics,
06099 Halle, Germany, frank.heyde@mathematik.uni-halle.de}, Birgit
Rudloff\thanks{Princeton University, ORFE, Princeton, NJ 08544, USA,
brudloff@princeton.edu}}

\usepackage{amsmath, amssymb}

\newtheorem{theorem}{Theorem}
\newtheorem{corollary}[theorem]{Corollary}
\newtheorem{remark}[theorem]{Remark}
\newtheorem{lemma}[theorem]{Lemma}
\newtheorem{definition}[theorem]{Definition}
\newtheorem{proposition}[theorem]{Proposition}

\newtheorem{example}[theorem]{Example}

\numberwithin{equation}{section} \numberwithin{theorem}{section}

\newcommand{\of}[1]{\ensuremath{\left( #1 \right)}}

\newcommand{\abs}[1]{\ensuremath{\left| #1 \right|}}
\newcommand{\cb}[1]{\ensuremath{ \left\{ #1 \right\} }}
\newcommand{\sqb}[1]{\ensuremath{ \left[ #1 \right] }}

\newcommand{\bs}{\backslash}

\newcommand{\vp}{\ensuremath{\varphi}}

\newcommand{\pend}{\hfill $\square$}

\newcommand{\R}{\mathrm{I\negthinspace R}}
\newcommand{\N}{\mathrm{I\negthinspace N}}
\newcommand{\One}{\mathrm{1\negthickspace I}}
\newcommand{\dom}{{\rm dom \,}}
\newcommand{\epi}{{\rm epi \,}}

\newcommand{\gr}{{\rm graph \,}}
\newcommand{\cl}{{\rm cl \,}}

\newcommand{\co}{{\rm co \,}}

\newcommand{\ri}{{\rm ri \,}}

\newcommand{\diag}{{\rm diag}}

\newcommand{\Int}{{\rm int\,}}

\begin{document}

\maketitle

\begin{abstract}
\noindent Set-valued risk measures on $L^p_d$ with $0 \leq p \leq \infty$ for conical
market models are defined, primal and dual representation results are given. The
collection of initial endowments which allow to super-hedge a multivariate claim are
shown to form the values of a set-valued sublinear (coherent) risk measure. Scalar risk
measures with multiple eligible assets also turn out to be a special case within the
set-valued framework.
\\[.2cm]
\\[.2cm]
\\[.2cm]
\end{abstract}

\section{Introduction}

The aim of this paper is to establish primal and dual representation results for
set-valued risk measures in conical markets, thus extending the results of
\cite{HamelHeyde10} to the case of random exchange rates at terminal time. The model
includes markets with bid-ask price spreads at initial and terminal time, generated for
example by proportional transaction costs. The model considered here is much more
realistic and better suited for applications than the ones used in \cite{HamelHeyde10}
and \cite{JouMedTou04} since, for example, those models cannot be linked with
no-arbitrage type results for conical market models. In this paper, we will provide the
link between set-valued risk measures and super-replication theorems for conical market
models, the latter established for example in \cite{Kabanov99}, \cite{Schachermayer04}.

Moreover, new features will turn up in the random exchange rate case, most notably that
risk measures may not be compatible with the chosen market model. This means that some
risk measures may not yield all risk compensating initial positions for some random
payoffs, since results of trading operations at initial and/or terminal time according to
the market may not be part of the acceptance set belonging to the risk measure. This
could be of advantage, for example, if there is no consensus about the market model, or
if one wishes to be very conservative.

From a mathematical point of view, we show that compatibility with the market at initial
time determines the image space of a set-valued risk measure, whereas compatibility with
the market at terminal time turns out to be a monotonicity property. This relationship
has not been observed before: In \cite{BurgertRueschendorf06} only scalar risk measures
for multivariate positions are considered without reference to a market model. In
\cite{JouMedTou04} and \cite{HamelHeyde10} (set-valued) risk measures for a static market
are defined  such that compatibility is automatic, whereas in \cite{CascosMolchanov07}
set-valued risk measures are investigated without reference to a market model.

Finally, we shall show that even scalar risk measures with multiple eligible assets as
proposed in \cite{ArtDelKochMed09} fit into our framework.

The almost natural occurrence of set-valued risk measures in market models with frictions
might also boost the theory of set-valued optimization problems. The authors do not know
any other application leading to a (primal) set-valued optimization problem: In the
context of the present paper, a problem like "minimize the (set-valued) risk of a
multivariate position subject to some constraints" makes perfectly sense from an
application point of view, and this opens the door to a all new research area.

The departing point is chosen similar to the one in \cite{ArtDelKochMed09}: We are given
a set of multivariate random variables which are specified according to the risk
tolerance of a regulator or supervisor (in the following: regulator). This set, the
regulator acceptance set, satisfies only minimal requirements. The regulator acceptance
set is, as usual, in one-to-one correspondence with a regulator risk measure. The
investor or financial agent (in the following: agent) may add particular features in
order to evaluate the risk of a multivariate outcome of her actions. For example, she
wants to (or has to) take a market model into account. This leads to augmented (agent)
acceptance sets which will have particular properties, and to corresponding risk
measures.

The augmentation procedure for the acceptance set as described in section
\ref{SubSecMarket} will shed new light on the interplay between risk specifications of a
regulator and the degrees of freedom for agents. For example, we think it is very
reasonable that a regulator need not be concerned about specific market models, but just
about tolerable outcomes. On the other hand, the market model already is a choice of the
agent. In this paper, we assume that the agent did choose a specific (conical) market
model. If she does not feel sure about the model, she at least could either do a
sensitivity analysis with respect to the market model, or just stick to the regulator
risk measure. The first possibility is beyond the scope of this paper, but it seems to be
a very interesting topic for future research. The second possibility is easily done using
the results below.

The tools used for modeling (convex) set-valued risk measures for multivariate random
variables and in particular, the duality formulas necessary to establish dual
representation results rely on the approach given in \cite{Hamel09}. It turns out that
this approach matches perfectly the needs of risk measures and superhedging theorems for
conical markets models. In particular, our dual variables, namely some simple set-valued
functions, and the conditions they have to satisfy have a meaning in terms of the
financial market model. We can hardly see how this goal could be achieved by sticking to
a vector-valued duality theory in which usually linear operators feature as dual
elements.

The main result of the paper is the dual representation theorem for convex set-valued
risk measures for conical market models (theorem \ref{ThmDualRep} below). It can be seen
as a duality result for a set-valued optimization problem. Indeed, if one starts with a
(closed convex) set $A$ of acceptable random variables the risk measure corresponding to
$A$ is given as an infimum in a set-valued sense which here is (the closure of) a union.
On the other hand, the duality result produces a representation of the set-valued (proper
closed convex) risk measure as a supremum of its set-valued affine minorants. The
supremum is taken in a complete lattice of sets with respect to "$\supseteq$", so it is
an intersection.

The paper is organized as follows. In section \ref{SecPrimal} we define regulator
acceptance sets, describe augmentation procedures, the construction of risk measures and
the one-to-one relationships between classes of set-valued risk measures and acceptance
sets. Sections \ref{SecDualVar} and \ref{SecDualRep} include the duality results. Section
\ref{SecExamples} is devoted to examples the most important of which is the link between
set-valued coherent risk measures and superhedging theorems for conical market models. We
also include set-valued versions of value at risk, the worst case risk measure and
average value at risk as well as a discussion of risk measures in a multiple eligible
asset market. In the appendix, some technical results are recalled or proven.

\section{From acceptance sets to risk measures}

\label{SecPrimal}

\subsection{Acceptance sets}

We are given a probability space $\of{\Omega, \mathcal{F}_T, P}$. By $L^0_d =
L^0_d\of{\Omega, \mathcal{F}_T, P}$ we denote the linear space of all $P$-measurable
functions $X \colon \Omega \to \R^d$, and by $L^p_d = L^p_d\of{\Omega, \mathcal{F}_T,
P}$, $0 < p \leq \infty$, the linear space of all such functions with $\int_\Omega
\abs{X\of{\omega}}^p \, dP < +\infty$ for $0 < p < \infty$, and $ess.sup_{\omega \in
\Omega} \abs{X\of{\omega}} < \infty$ for $p = \infty$. In all cases, $\abs{\cdot}$ stands
for an arbitrary, but fixed norm on $\R^d$, and the usual identification of functions
differing only on sets of $P$-measure zero is assumed. As usual, we write
\[
\of{L^p_d}_+ = \cb{X \in L^p_d \colon X \in \R^d_+ \; P-\mbox{a.s.}}
\]
for the closed convex cone of $\R^d$-valued random variables with $P$-almost surely
non-negative components. An element $X \in L^p_d$ has components $X_1, \ldots, X_d$ in
$L^p = L^p_1$. The symbol $\One$ denotes the random variable in $L^p_1$ which has
$P$-almost surely the value $1$.

In this paper, for  $X \in L^p_d$, the values of $X_i$, $1 \leq i \leq d$, are understood
as the number of units of asset $i$ an agent holds at terminal time $T$. Thus, we follow
Kabanov's idea \cite{Kabanov99} in assuming that a portfolio is represented in "physical
units" of the traded assets (instead of its value in a fixed currency/num\'{e}raire).

Following \cite{ArtDelKochMed09} we consider an acceptance set given by a regulator as
the object with "conceptual primacy [...] in risk measurement".

\begin{definition}
\label{DefAccSet} An {\em acceptance set} is a subset $A \subseteq L^p_d$ with $0 \leq p
\leq \infty$ satisfying $\R^d\One \cap A \neq \emptyset$, $\R^d\One \cap \of{L^p_d\bs A}
\neq \emptyset$ and $A + \of{L^p_d}_+ \subseteq A$.
\end{definition}

We consider the properties in definition \ref{DefAccSet} as the minimal requirements to
be asked of an acceptance set. We see them as rational in the sense that every regulator
would agree upon them. The first two simply mean that there is a deterministic portfolio
which is accepted by the regulator, but the regulator does not accept all deterministic
portfolios. The last two properties imply a weak condition of the boundedness-from-below
type: There is $x \in \R^d$ such that $\of{x\One - \of{L^p_d}_+} \cap A = \emptyset$.
Indeed, otherwise for each $x \in \R^d$ there would exist $X \in \of{L^p_d}_+$ such that
$x\One - X \in A$, hence $x\One \in X + A \subseteq A$. This would be $\R^d\One \subseteq
A$ contradicting $\R^d\One \cap \of{L^p_d\bs A} \neq \emptyset$. Compare
\cite{FoellmerSchied04}, (4.3), (4.4) for similar requirements in the univariate case.

Note that in \cite{ArtDelKochMed09} (p. 3, (iii)) a subset of $L^0_1$ is called an
acceptance set if its elements, interpreted as "future values of financial positions",
are "in line with the risk tolerance of the supervisor", but no mathematical assumptions
are required. We deviate a little from this approach since otherwise the term "acceptance
set" would just be tantamount with "subset of $L^0_d$".

\subsection{Eligible portfolios}

A regulator usually asks for a deposit, to be made at initial time, to compensate for the
risk of an investment. This deposit could be given, for instance, in cash of a specific
currency, cash of several currencies, units of other num\'{e}raires, or even positions
with fixed proportions of several of such assets. The set of such positions spans a
linear space $M$, the space of eligible portfolios. Without loss of generality, we assume
$M \subseteq \R^d$, i.e. all eligible assets are included in the set of the "traded"
assets. The following two conditions describe the relationship between the space of
eligible portfolios and an acceptance set:
\\
(A1a) $M\One \cap A \neq \emptyset$ and
\\
(A1b) $M\One \cap \of{L^p_d \bs A} \neq \emptyset$.

\medskip (A1a) says that there is an eligible portfolio which is acceptable at terminal
time. (A1b) says that not every eligible portfolio is acceptable at terminal time. Of
course, (A1a), (A1b) are stronger than the first two requirements in definition
\ref{DefAccSet}.

\subsection{Risk measures}

A risk measure will turn out to be a function which maps multivariate random variables
into the power set $\mathcal{P}\of{M}$ of the space $M$ of eligible portfolios (including
the empty set $\emptyset$). We associate with a function $R \colon L^p_d \to
\mathcal{P}\of{M}$, as usual, its graph defined by
\[
\gr R = \cb{\of{X, u} \in L^p_d \times M \colon u \in R\of{X}}.
\]

\begin{definition}
\label{DefMTransMono} A function $R \colon L^p_d \to \mathcal{P}\of{M}$ is called {\em
translative in $M$} or just {\em $M$-translative} iff
\begin{equation}
\label{EqTrans} \forall X \in L^p_d, \; \forall u \in M \colon
    R\of{X + u\One} =  R\of{X} - u.
\end{equation}
Let $B \subseteq L^p_d$. A function $R \colon L^p_d \to \mathcal{P}\of{M}$ is called {\em
$B$-monotone} iff
\[
X^2 - X^1 \in B \quad \Rightarrow \quad R\of{X^2} \supseteq R\of{X^1}.
\]
\end{definition}

In the following definition, we identify risk measures among all functions into
$\mathcal{P}\of{M}$.

\begin{definition}
\label{DefRiskMea} A {\em risk measure} is a function $R \colon L^p_d \to
\mathcal{P}\of{M}$ with $0 \leq p \leq \infty$ which is $M$-translative and
$\of{L^p_d}_+$-monotone. A risk measure is said to be finite at zero if it satisfies
\\
(R1a) $R\of{0} \neq \emptyset$ and
\\
(R1b) $R\of{0} \neq M$.
\end{definition}

The value $R\of{X}$ of a risk measure $R$ is understood to include the eligible portfolio
vectors which compensate for the risk of $X$.

The interpretation of (R1a) is, of course, that there is an eligible portfolio at initial
time which compensates for the risk of the zero payoff at terminal time. (R1b) just means
that not all eligible portfolios compensate for the risk of the zero payoff.

The interpretation of $M$-translativity is straightforward if one recalls the scalar
case, see \cite{ArtDelEbeHea99}, \cite{ArtDelKochMed09}, \cite{FoellmerSchied04}.

The monotonicity condition says that a random portfolio vector which is componentwise not
less than another in (almost) all scenarios should admit more risk compensating eligible
portfolios. Therefore, the partial order $\supseteq$ is the relation of choice in the
image space $\mathcal{P}\of{M}$.

Let a set $A \subseteq L^p_d$ be given. By
\begin{equation}
\label{EqRA} R_A\of{X} = \cb{u \in M \colon X + u\One \in A}
\end{equation}
a function $R_A \colon L^p_d \to \mathcal{P}\of{M}$ is defined. If $A$ is an acceptance
set, then the interpretation of \eqref{EqRA} is, of course, that $R_A\of{X}$ includes all
eligible portfolios which, when added to $X$, compensate for the risk of $X$, i.e. lead
to an overall position in $A$.

Conversely, we associate with a function $R \colon L^p_d \to \mathcal{P}\of{M}$ the set
\begin{equation}
\label{EqAR} A_R = \cb{X \in L^p_d \colon 0 \in R\of{X}}.
\end{equation}
If $R$ is a risk measure, then $A_R$ includes those positions $X$ which have zero among
its risk compensating eligible portfolios, i.e. a position $X$ is acceptable in terms of
the risk measure $R$ if it can be made acceptable without additional initial endowment.

As in the scalar case, there is a one-to-one correspondence between families of
$M$-translative functions and families of subsets of $L^p_d$ via \eqref{EqRA},
\eqref{EqAR}. The following simple result provides the basic relationship.

\begin{proposition}
\label{PropOneToOneB} Let $A \subseteq L^p_d$ be given. Then, the function $R_A$ is
$M$-translative and $A = A_{R_A}$. Conversely, let $R \colon L^p_d \to \mathcal{P}\of{M}$
be $M$-translative. Then $R = R_{A_R}$.
\end{proposition}

{\sc Proof.} From (\ref{EqRA}), (\ref{EqAR}) we get
\[
A_{R_A} = \cb{X \in L^p_d \colon 0 \in R_A\of{X}}
    = \cb{X \in L^p_d \colon 0 \in \cb{u \in M \colon X + u\One \in A}} = A
\]
as desired. A direct calculation using \eqref{EqRA}, \eqref{EqAR} and \eqref{EqTrans}
yields
\begin{eqnarray*}
R_{A_R}\of{X} & = & \cb{u \in M \colon X + u\One \in A_R} \\
    & = & \cb{u \in M \colon  0 \in R\of{X + u\One}} = R\of{X}.
\end{eqnarray*}
This completes the proof. \pend

\begin{proposition}
\label{PropOneToOneAR} Let $A \subseteq L^p_d$ be an acceptance set. Then, the function
$R_A$ is a risk measure. If $A$ satisfies (A1a), (A1b), then $R_A$ is finite at zero.
Conversely, let $R \colon L^p_d \to \mathcal{P}\of{M}$ be a risk measure. Then, $A_R$ is
an acceptance set. If $R$ is finite at zero, then $A_R$ satisfies (A1a), (A1b).
\end{proposition}

{\sc Proof.} This follows from proposition \ref{PropOneToOneB} and
\ref{PropPrimalRepresProp} (a) (with $B = \of{L^p_d}_+$), (b), (c) in the appendix. \pend

\medskip Additional properties may (and will) be required for acceptance
sets and risk measures. Most importantly, conditions for the compatibility of the
acceptance set with the market model are given in the following section.

\subsection{Market-compatibility}

\label{SubSecMarket}

So far, acceptance sets and risk measures have nothing to do with the financial market.
This is on purpose: The regulator should not be concerned about the question what market
model "describes the reality". Moreover, we see the market model as a choice of the
agent: Agents may not agree upon the model.

Once a market model is chosen, one can describe consequences for the acceptance set and
the corresponding risk measure. Here, we consider a one-period, conical market model. It
can be seen as the one-period special case of the conical models considered in
\cite{PennanenPenner08R}. Such a model may occur if proportional transaction costs are
present (see \cite{Kabanov99} and also \cite{JouMedTou04}), or a bid-ask price spread is
modeled directly (see e.g. \cite{Schachermayer04}).

At initial time, a closed convex cone $K_I \subseteq \R^d$ with $\R^d_+ \subseteq K_I
\neq \R^d$ is given which models the proportional frictions between the assets according
to the geometric model introduced in \cite{Kabanov99}. This cone is called solvency cone
since it includes precisely those (deterministic) portfolios which can be exchanged at
initial time into portfolios with only non-negative components. The part of the cone
$K_I$ that is relevant for the space $M$ of eligible portfolios is $K_I^M = K_I \cap M$,
also a closed convex cone. It is assumed throughout the paper that $M \cap \R^d_+ \neq
\cb{0}$. This implies in particular $K^M_I \neq \cb{0}$ since $\R^d_+ \subseteq K_I$.

At terminal time, the market is described by means of a measurable mapping $\omega
\mapsto K_T\of{\omega}$, the solvency cone mapping, with $K_T\of{\omega} \subseteq \R^d$
a closed convex cone such that $\R^d_+ \subseteq K_T\of{\omega} \neq \R^d$ for all
$\omega \in \Omega$. For $0 \leq p \leq \infty$, the set
\begin{equation}
\label{EqConeC} L^p_d\of{K_T} = \cb{X \in L^p_d \colon
    P\of{\cb{\omega \in \Omega \colon X\of{\omega} \in K_T\of{\omega}}} = 1}
\end{equation}
is a closed convex cone in $L^p_d$ generating a reflexive transitive relation for
$\R^d$-valued random variables. We recall the following definition, see for example
\cite{AubinFrankowska90}.

\begin{definition}
A set-valued function $K \colon \Omega \to \mathcal{P}\of{\R^d}$ is called measurable iff
for each open set $B \subseteq \R^d$ the set $\cb{\omega \in \Omega \colon K\of{\omega}
\cap B \neq \emptyset}$ is an element of $\mathcal{F}_T$.
\end{definition}

In contrast to \cite{HamelHeyde10}, we allow for the additional randomness expressed by
the measurable set-valued function $K_T$: It reflects the fact that the future
transaction costs and/or exchange rates are not known at initial time. In
\cite{HamelHeyde10}, $K= K_I \equiv K_T$ is assumed as in \cite{JouMedTou04}.

According to \eqref{EqRA}, if a financial agent is given an acceptance set and a final
payoff $X \in L^p_d$, she is supposed to look for all eligible portfolios $u \in M$ which
make the overall position $X + u\One$ acceptable. Since trading is possible at initial
and terminal time, two situations may occur.

First, the agent might be interested to know about all eligible portfolios which can be
exchanged into an eligible portfolio which in turn makes a given payoff $X \in L^p_d$
acceptable. Suppose $u \in M$ is such that $u' \in u - K^M_I$ makes $X$ acceptable, that
is, the investor can exchange $u$ into $u'$ at initial time (and stay in $M$) in order to
get a risk compensating eligible portfolio for $X$. Then $X + u\One \in A + K^M_I\One$,
that is, the agent looks for
\[
R_{A + K^M_I\One}\of{X} = \cb{u \in M \colon X + u\One \in A + K^M_I\One}
    = R_A\of{X} + K^M_I.
\]
(the reader easily verifies the last equation). From the point of view of the regulator
it does not matter if the agent starts with initial endowment $u$ or $u'$: If $A$ is the
acceptance set given by the regulator and $X + u\One \not\in A$, then the agent has to
exchange $u$ into $u'$ and give $u'$ as deposit in order to make $X$ acceptable. For the
agent, however, it could make a difference since a transaction might not be desirable for
her. On the other hand, it clearly is of advantage for the agent to know all initial
endowments which -- maybe after a transaction at initial time -- admit to compensate for
the risk of $X$.

Secondly, the agent might be interested to know what risks she could cover with a given
available eligible portfolio $u \in M$. Of course, every $X \in L^p_d$ such that $X +
u\One \in A$. Moreover, if $X + u\One \not\in A$, but she can exchange $X$ for $X'$ at
terminal time such that $X' + u\One \in A$, it is reasonable for her to consider $u$ as a
risk compensating portfolio for $X$, too: Indeed, there is no additional initial
endowment necessary, but just an exchange from $X$ to $X'$ at time $T$. Since in this
case $X' \in X - K_T$ a.s. and $X' + u\One \in A$ we have $X + u\One \in A +
L^p_d\of{K_T}$. Thus, the agent looks for
\[
R_{A + L^p_d\of{K_T}}\of{X} = \cb{u \in M \colon X + u\One \in A + L^p_d\of{K_T}}.
\]
Again, from the point of view of the regulator it does not matter how the agent ends up
with a position in $A$. From the point of view of the agent it could make a difference:
Not only that an exchange might not be desirable, the chosen market model could be wrong.
In the latter case, the required exchange at terminal time could be impossible leaving
the agent with a non-acceptable position. This certainly is a motivation for the agent to
look at both of $R_A$ and $R_{A + L^p_d\of{K_T}}$.

The above considerations justify the following definition.

\begin{definition}
\label{DefMarketComp} An acceptance set $A$ is called $K_T$-compatible iff $A +
L^p_d\of{K_T} \subseteq A$, and it is called $K_I$-compatible iff $A + K^M_I\One
\subseteq A$.

A risk measure is called $K_T$-compatible iff it is $L^p_d\of{K_T}$-monotone, and it is
called $K_I$-compatible iff it maps into the set $\mathcal{P}^M_K = \cb{D \subseteq M
\colon D = D + K^M_I}$.

An acceptance set or a risk measure is called market-compatible iff it is $K_I$- and
$K_T$-compatible.
\end{definition}

Trivially, if $A$ is an acceptance set, $A + L^p_d\of{K_T}$ is $K_T$-compatible, $A +
K^M_I\One$ is $K_I$-compatible, and $A + L^p_d\of{K_T} + K^M_I\One$ is market-compatible,
and $A$ is market-compatible if and only if $A + L^p_d\of{K_T} + K^M_I\One = A$. Of
course it is of interest (and done below) to establish conditions under which $R_A\of{X}=
R_A\of{X} + K^M_I$ and/or $R_A = R_{A + L^p_d\of{K_T}}$, respectively. The reader may
note that the question wether or not a risk measure is market-compatible is a new feature
in the multi-variate case which does not occur (or is trivial) if scalar risk measures
for univariate positions are under consideration.

\begin{proposition}
\label{PropMarketComp} Let $A \subseteq L^p_d$ be a $K_T$-compatible ($K_I$-compatible,
market-compatible) acceptance set. Then, $R_A$ is a $K_T$-compatible ($K_I$-compatible,
market-compatible) risk measure.

Conversely, let $R \colon L^p_d \to \mathcal{P}\of{M}$ be a $K_T$-compatible
($K_I$-compatible, market-compatible) risk measure. Then, $A_R$ is a $K_T$-compatible
($K_I$-compatible, market-compatible) acceptance set.
\end{proposition}

{\sc Proof.} Follows from proposition \ref{PropPrimalRepresProp} (a), (d). \pend

\medskip The following simple example shows that not every "intuitive" acceptance set is market-compatible.

\begin{example}
\label{ExWC2D} Let $d = m = 2$ and $K_I = K^M_I = \cb{x \in \R^2 \colon x_2 \geq -2x_1,
\; x_2 \geq 0}$. For all $\omega \in \Omega$ let $K_T = \cb{x \in \R^2 \colon x_2 \geq
-\frac{1}{2}x_1, \; x_1 \geq 0}$. The point $u = \of{-1, 4}$ is in $K_I$, but not in
$K_T$. Let us for the moment accept positions in $A = L^p_2\of{K_T}$. Then $X = 0 \in
L^p_2$ is certainly an acceptable position. However, the position $X + u\One = u\One$ is
not acceptable despite the fact that $u$ can be transferred into a "strictly" positive
position at initial time, that is $A$ is not $K_I$-compatible. On the other hand, of
course $A$ is $K_T$-compatible.
\end{example}

As in the previous example, the "intuitive" worst case acceptance set $L^p_d\of{K_T}$ is
the standard example of a set which is $K_T$-compatible, but not market-compatible in
general. See section \ref{SecWCRF} for more details.

\begin{remark}
\label{RemPriceZero} The set $-L^p_d\of{K_T} - K_I\One$ is the set of all terminal
positions which can be generated with zero initial costs, that is starting with the zero
portfolio at initial time. A pair $\of{u', X'} \in \R^d \times L^p_d$ satisfying $u' \in
-K_I$ and $X' \in u'\One - K_T$ almost surely is called a self-financing portfolio
process for the one-period market $\of{K_I, K_T}$ (see \cite{Schachermayer04}). Thus, $u
\in M$ makes $X \in L^p_d$ acceptable with respect to the market-compatible acceptance
set $A + L^p_d\of{K_T} + K^M_I\One$ if and only if there is a self-financing portfolio
process $\of{u', X'}$ such that $u' \in M$ and $X + X' + u\One \in A$.
\end{remark}

\begin{remark} \label{RemKMIAug} The set $A+ K_I^M\One$ satisfies (A1b) if and only if  $A  \subseteq L^0_d$
satisfies a certain boundedness from below condition with respect to the cone $K^M_I$.
Indeed,
\begin{align*}
A + K_I^M\One \text{ satisfies (A1b) }
& \Leftrightarrow \exists u \in M \colon u\One \not\in A + K_I^M\One\\
& \Leftrightarrow \exists u \in M \; \forall k \in K^M_I \colon \of{u-k}\One \not\in A\\
& \Leftrightarrow \exists u \in M \colon \of{u-K^M_I}\One \cap A = \emptyset.
\end{align*}
With a similar argument, one may verify that $A \subseteq L^0_d$ satisfies (A1a) if and
only if $A + K_I^M\One$ does.
\end{remark}

\subsection{Diversification and convexity issues}

It is widely acknowledged that convexity is a useful property of acceptance sets and risk
measures. For example, the criticism against the use of value at risk as a risk measure
is based on the lack of convexity.

While the definition of convexity for an acceptance set $A \subseteq L^p_d$, being a
subset of a linear space, is the usual one, we use the following definitions for
functions $R \colon L^p_d \to \mathcal{P}\of{M}$. The reader who is familiar with
\cite{HamelHeyde10} may observe that this approach deviates from the one in the quoted
paper where only market-compatible risk measures have been considered.

\begin{definition}
\label{DefPropFunctions} A function $R \colon L^p_d \to \mathcal{P}\of{M}$ is called
convex iff
\[
\forall X, X' \in L^p_d, \; \forall t \in \of{0,1} \colon
    R\of{tX + \of{1-t}X'} \supseteq t R\of{X} + \of{1-t}R\of{X'}.
\]
It is called positively homogeneous iff
\[
\forall X \in L^p_d, \; \forall t \in \of{0,1} \colon
    R\of{tX} \supseteq t R\of{X},
\]
and it is called subadditive iff
\[
\forall X, X' \in L^p_d \colon
    R\of{X + X'} \supseteq R\of{X} + R\of{X'}.
\]
A positively homogeneous and subadditive function is called sublinear.
\end{definition}

The above definition of convexity just means that the "mixed" position $tX + \of{1-t}X'$
admits at least the risk compensating eligible portfolios which could be obtained by
mixing the ones for $X$ and $X'$. Alternative characterizations of these properties in
terms of the graph of $R$ can be found in the appendix, section \ref{SecAppConvex}.
Moreover, the values of a convex set-valued function $R$ are convex (choose $X = X'$ in
the above definition), and $R\of{0}$ is a cone if $R$ is positively homogeneous. If $R$
is sublinear, then the convex cone $R\of{0}$ is included in the recession cone of
$R\of{X}$ for each $X$ since $R\of{X+0} \supseteq R\of{X} + R\of{0}$ for each $X \in
L^p_d$.

\begin{proposition}
\label{PropConvexity} Let $A \subseteq L^p_d$ be an acceptance set which is convex
(closed under addition, a cone). Then, $R_A$ is a convex (subadditive, positively
homogeneous) risk measure. Conversely, let $R \colon L^p_d \to \mathcal{P}\of{M}$ be a
convex (subadditive, positively homogeneous) risk measure. Then, $A_R$ is an acceptance
set which is convex (closed under addition, a cone).
\end{proposition}

{\sc Proof.} Follows from proposition \ref{PropPrimalRepresProp} (e), (f), (g). \pend

\medskip As usual, a sublinear risk measure is called coherent. We immediately obtain the
one-to-one correspondence between coherent risk measures (and being finite at zero) and
acceptance set which are convex cones (and satisfying (A1a), (A1b)).

\begin{remark}
\label{RemConvexCompatible} If the acceptance set $A$ is convex (a cone), then the
market-compatible augmented acceptance set $A + L^p_d\of{K_T} + K^M_I\One$ again is
convex (a cone).
\end{remark}

\subsection{Closedness}

Closedness assumptions are inevitable for convex duality results. Given an acceptance set
$A$, the corresponding risk measure $R_A$ may have two different closedness properties.
First, it may have closed values. Secondly, its graph may be closed. These two properties
correspond to two different properties of the acceptance set $A$. While the one
corresponding to closedness of the graph is closedness of $A$ in $L^p_d$, the closedness
property of $A$ which corresponds to closed values of $R_A$ is more involved. In the
following definition, we use the symbol $u^k \stackrel{M}{\longrightarrow} 0$ in order to
denote a sequence $\cb{u^k}_{k \in \N} \subset M$ with $\lim_{k \to \infty} u^k = 0$.

\begin{definition}
\label{DefRadCloSet} A set $A \subseteq L^p_d$ is called {\em directionally closed} in
$M$ iff $X \in L^p_d$, $u^k \stackrel{M}{\longrightarrow} 0$ and $X + u^k\One \in A$ for
all $k \in \N$ imply $X \in A$. For an arbitrary set $A \subseteq L^p_d$, the set
\[
\cl_M A = \cb{X \in L^p_d \colon \exists u^k \stackrel{M}{\longrightarrow} 0 \colon
    \forall k \in \N \colon X + u^k\One \in A}
\]
is called the directional closure of $A$ in $M$.
\end{definition}

Of course, a closed set $A$ is directionally closed with respect to each linear subspace
$M \subseteq \R^d$. See the appendix for more information about this property. Note also
that a set-valued function $R$ with a closed graph has closed values.

\begin{proposition}
\label{PropClosedness} Let $A \subseteq L^p_d$ be an acceptance set which is
directionally closed with respect to $M$ (closed in $L^p_d$). Then, $R_A$ is a risk
measure with closed values (a closed graph). Conversely, let $R \colon L^p_d \to
\mathcal{P}\of{M}$ be a risk measure with closed values (a closed graph). Then, $A_R$ is
an acceptance set which is directionally closed with respect to $M$ (closed in $L^p_d$).
\end{proposition}

{\sc Proof.} Follows from proposition \ref{PropPrimalRepresProp} (i), (k). \pend

\begin{remark}
\label{RemClosedCompatible} If the acceptance set $A$ is closed (directionally closed in
$M$), then the market-compatible augmented acceptance set $A + L^p_d\of{K_T} + K^M_I\One$
need not to be closed (directionally closed in $M$) in general.
\end{remark}

\begin{remark}
\label{RemClosedExtension} Lemma \ref{LemDirClosure2} (see appendix) makes the following
construction possible. If an acceptance set is not directionally closed in $M$ (or one
just does not know), then the function
\[
X \mapsto \cl\cb{u \in M \colon X + u\One \in A}
\]
coincides with $R_{\cl_M A}$. Since $\cl_M$ is a hull operator (see lemma
\ref{LemDirClosure1}), for each acceptance set $B \subseteq L^p_d$ with $A \subseteq B$
which is directionally closed in $M$ we have
\[
\forall X \in L^p_d \colon R_B\of{X} \supseteq R_{\cl_M A}\of{X}.
\]

Similarly, if $A$ is not closed (or one just does not know), then the graph of the
function
\[
X \mapsto \cb{u \in M \colon X + u\One \in \cl A}
\]
coincides with $\cl\of{\gr R_A}$ (and is of course $\gr R_{\cl A}$). For each
$M$-translative function $R$ with a closed graph including the graph of $R_A$ we have
\[
\forall X \in L^p_d \colon R\of{X} \supseteq R_{\cl A}\of{X}.
\]
\end{remark}

\section{Dual variables}

\label{SecDualVar}

In the next two sections, we will give dual representation results. Therefore, we shall
assume $1 \leq p \leq \infty$ in the following. In case $p = \infty$ we consider the dual
pair $\of{L^\infty_d, L^1_d}$ with $\sigma\of{L^\infty_d, L^1_d}$-topology on
$L^\infty_d$. In all other cases, $L^p_d$ is paired with $L^q_d$ for $\frac{1}{p} +
\frac{1}{q} = 1$ and their respective norm topologies.

By $K^+_I$ and $\of{K^M_I}^+$ we denote the positive dual cones of the cones $K_I$ in
$\R^d$ and $K^M_I$ in $M$, respectively. Thus,
\[
\of{K^M_I}^+ = \cb{v \in M \colon \forall u \in K^M_I \colon v^Tu \geq 0} \subseteq M.
\]
Note that $\of{K^M_I}^+ = \of{K^+_I + M^\perp}\cap M$ with $M^\perp = \cb{v \in \R^d
\colon \forall u \in M \colon v^Tu = 0}$ since $K^+_I + M^\perp$ is the dual cone of
$K_I^M$ in $\R^d$. The reader should be aware that both $\of{K^M_I}^+$ and $K^+_I +
M^\perp$ are dual cones of $K_I^M$, the first one in $M$, the second one in $\R^d$. We
will also use $G\of{v} = \cb{x \in \R^d \colon 0 \leq v^Tx}$ for $v \in \R^d$ being a
subset of $\R^d$, not necessarily one of $M$.

By $K^+_T$ we denote the set--valued mapping $\omega \mapsto \sqb{K_T\of{\omega}}^+$, and
$\sqb{L^p_d\of{K_T}}^+ \subseteq L^q_d$ denotes the dual cone of $L^p_d\of{K_T}$.

\begin{lemma}
\[
\sqb{L^p_d\of{K_T}}^+ = L^q_d\of{K^+_T} =
    \cb{Y \in L^q_d \colon Y\of{\omega} \in K^+_T\of{\omega} \; P-a.s.}.
\]
\end{lemma}

{\sc Proof.} ''$\supseteq$'': Let $Y\in L^q_d\of{K^+_T}$ and $X\in L^p_d\of{K_T}$. Then,
the relationships $X \in K_T$, $Y \in K^+_T$ and $X^T Y \geq 0$ hold $P$--almost surly.
Consequently, $E\sqb{X^TY} \geq 0$.

''$\subseteq$'': Assume $Y\in \sqb{L^p_d\of{K^+_T}}^+$ but $Y\not\in L^q_d\of{K^+_T}$.
Then, there is a set $A\in \mathcal{F}$ with $P\of{A}>0$ such that $Y\of{\omega}\not\in
K^+_T\of{\omega}$ for all $\omega \in A$. The function $f \colon \Omega \times \R^d \to
\R$ defined by $f\of{\omega, x} = x^TY\of{\omega}$ is a Caratheodory function. The
set-valued map $F \colon \Omega \to \mathcal{P}\of{\R^d}$ defined by $F\of{\omega} =
K_T\of{\omega} \cap B_1\of{0}$ is measurable and compact--valued. Since, for each $\omega
\in \Omega$, $f\of{\omega, \cdot}$ is continuous and $F\of{\omega}$ is compact,
$\Phi\of{\omega} = \cb{x \in F\of{\omega} \colon f\of{\omega, x} = \inf_{y \in
F\of{\omega}} f\of{\omega,y}}$ is nonempty. Moreover, $\Phi\of{\omega}$ is closed. By
\cite[theorem 8.2.11]{AubinFrankowska90}, $\Phi \colon \Omega \to \mathcal{P}\of{\R^d}$
is a measurable set-valued map which has a measurable selection $X$ by \cite[theorem
8.1.3]{AubinFrankowska90}. Since, for all $\omega \in \Omega$,
\[
X\of{\omega}^TY\of{\omega} = \inf_{y \in F\of{\omega}} y^T Y\of{\omega} \leq 0
\]
because of $0\in F\of{\omega}$ and, for all $\omega\in A$, $X\of{\omega}^TY\of{\omega}
 < 0$ because of $Y\of{\omega} \not\in K^+_T\of{\omega}$ we have $E\sqb{X^TY} < 0$,
a contradiction to $Y\in \sqb{L^p_d\of{K_T}}^+$. \pend

\medskip The following definition introduces a class of set-valued functions which will
be used as a substitute for continuous linear functionals on $L^p_d$, $p \geq 1$. These
functions depend on an additional dual variable which reflects the order relation in the
image spaces $\R^d$ and $M$, generated by the cones $K_I$ and $K^M_I$, respectively.

\begin{definition}
\label{DefSetExpect} Take $Y \in L^q_d$ and $v \in M$. Define a function $F^M_{\of{Y, v}}
\colon L^p_d \to \mathcal{P}\of{M}$ by
\begin{equation}
\label{DefSetExpectation} F^M_{\of{Y, v}}\sqb{X} = \cb{u \in M \colon E\sqb{X^T Y} \leq
v^T u}.
\end{equation}
\end{definition}

The following proposition collects elementary properties of these functions for future
reference. Compare also proposition 4.1 in \cite{HamelHeyde10} and proposition 6 in
\cite{Hamel09}.

\begin{proposition}
\label{PropExpectationRM} Let $Y \in L^q_d$ and $v \in M$. Then, the function $R \colon
L^p_d \to \mathcal{P}\of{M}$ defined by $R\of{X} = F^M_{\of{Y, v}}\sqb{-X}$
\\
(a) is additive and positively homogeneous with $F^M_{\of{Y, v}}\sqb{0} = G\of{v}\cap M =
\cb{x \in M \colon 0 \leq v^Tx}$;
\\
(b) has a closed graph, and hence closed values, namely closed half spaces;
\\
(c) is finite at zero if and only if it is finite everywhere if and only if $v \in
M\bs\cb{0}$; moreover, $R\of{X} \in \cb{M, \emptyset}$ for each $X \in L^p_d$ if and only
if $v=0$;
\\
(d) is $M$-translative if and only if $v \in E\sqb{Y} + M^\perp$;
\\
(e) is $K_T$-compatible if and only if $Y \in L^q_d\of{K^+_T}$;
\\
(f) is $K_I$-compatible if and only if $v \in \of{K^M_I}^+\bs\cb{0}$
\\
(g) has the acceptance set $A_R = \cb{X \in L^p_d \colon 0 \leq E\sqb{X^T Y}}$.
\end{proposition}

{\sc Proof.} (a) Can be checked directly, see proposition 4.1 in \cite{HamelHeyde10}. (b)
The graph of $R$ is the set $\cb{\of{X, u} \in L^p_d \times M \colon E\sqb{X^T Y} \leq
v^T u}$ which is a closed half space in $L^p_d \times M$ since $X \mapsto E\sqb{X^T Y}$
is a continuous linear functional on $L^p_d$ (also with respect to the
$\sigma\of{L^\infty_d, L^1_d}$-topology if $p=\infty$). (c) Easily checked. (d) Take $X
\in L^p_d$, $u \in M$. Then
\begin{align*}
F^M_{\of{Y, v}}\sqb{-X - u\One} & =
     \cb{z \in M \colon E\sqb{\of{X + u\One}^T Y}  + v^T z \geq 0} \\
     & = \cb{z \in M \colon E\sqb{X^T Y} + E\sqb{Y}^T u + v^T z \geq 0}\\
     & = \cb{z + u \in M \colon E\sqb{X^T Y} + \of{E\sqb{Y} - v}^Tu + v^T \of{z + u} \geq 0} - u \\
     & = F^M_{\of{Y, v}}\sqb{-X} - u.
\end{align*}
The last equation in this chain is true if and only if $E\sqb{Y} - v \in M^\perp$. (e)
Take $X^1, X^2 \in L^p_d$ such that $X^2 - X^1 \in L^p_d\of{K_T}$, and $u \in F^M_{\of{Y,
v}}\sqb{-X^1}$. Then
\[
E\sqb{\of{-X^2}^T Y} + E\sqb{\of{X^2 - X^1}^T Y} \leq v^T u,
\]
and since $Y \in L^q_d\of{K^+_T}$ we have $E\sqb{\of{X^2 - X^1}^T Y} \geq 0$. This proves
$u \in F^M_{\of{Y, v}}\sqb{-X^2}$, hence $R$ is $K_T$-compatible. Conversely, if $Y
\not\in L^q_d\of{K^+_T}$ there is $X \in L^p_d\of{K_T}$ such that $E\sqb{X^T Y} < 0$.
From $K_T$-compatibility with $X^1 = -X$, $X^2 = 0$ we get
\[
F^M_{\of{Y, v}}\sqb{X} \subseteq G\of{v} \cap M
\]
which is
\[
\cb{u \in M \colon E\sqb{X^T Y} \leq v^T u} \subseteq \cb{u \in M \colon 0 \leq v^T u}.
\]
Since $E\sqb{X^T Y} < 0$ there is a $u \in M$ belonging to the left hand side which is
not an element of the right hand side. This contradicts $Y \not\in L^q_d\of{K^+_T}$. (f)
This simply follows since $v^T\of{u} \leq v^T\of{u + k}$ for all $k \in K^M_I$ if and
only if $v \in \of{K^M_I}^+$. (g) Obvious. \pend

\begin{remark}
The functions $X \mapsto F^M_{\of{Y, v}}\sqb{X}, F^M_{\of{Y, v}}\sqb{-X}$ map into the
collection
\[
\mathbb{G}_M = \cb{D \subseteq \R^d \colon D = \cl\co\of{D + K^M_I}}
\]
iff $v \in \of{K^M_I}^+$.
\end{remark}

The next result admits a change of variables from vector densities $Y$ to vector
probability measures $Q$. This allows a formulation of the dual representation result in
terms of probability measures as it is common in the scalar case. In contrast,
super-hedging and no-arbitrage type results for conical market models are usually
formulated in terms of consistent pricing processes, see e.g. \cite{Schachermayer04},
\cite{PennanenPenner08R} and also \cite{JouMedTou04}. We will discuss the relationship in
more detail below.

In the following, $\diag\of{w}$ with $w \in \R^d$ denotes the diagonal matrix with the
components of $w$ as entries in its main diagonal and zero elsewhere. Moreover,
$\mathcal{M}^P_{1,d} = \mathcal{M}^P_{1,d}\of{\Omega, \mathcal{F}_T}$ denotes the set of
all vector probability measures with components being absolutely continuous with respect
to $P$, i.e. $Q_i \colon \mathcal{F}_T \to \sqb{0,1}$ is a probability measure on
$\of{\Omega,\mathcal{F}_T}$ such that $\frac{dQ_i}{dP} \in L^1$ for $i = 1, \ldots, d$.

\begin{lemma}
\label{LemDualTransform} (i) Let $Y \in L^q_d\of{K^+_T}$, $v \in \of{E\sqb{Y} + M^\perp}
\cap \of{K^M_I}^+\bs\cb{0}$. Then there are $Q \in \mathcal{M}^P_{1,d}$, $w \in K^+_I\bs
M^\perp + M^\perp$ such that $\diag\of{w}\frac{dQ}{dP} \in L^q_d\of{K^+_T}$ and
$F^M_{\of{Y, v}} = \widetilde{F}^M_{\of{Q, w}}$ with
\begin{equation}
\label{EqTransform}
 \widetilde{F}^M_{\of{Q, w}}\sqb{X} = \cb{u \in M \colon w^TE^Q\sqb{X}
 \leq w^T u} = \of{E^Q\sqb{X} + G\of{w}} \cap M.
\end{equation}
(ii) Vice versa, if $Q \in \mathcal{M}^P_{1,d}$, $w \in K^+_I\bs M^\perp + M^\perp$ such
that $\diag\of{w}\frac{dQ}{dP} \in L^q_d\of{K^+_T}$ then there are $Y \in
L^q_d\of{K^+_T}$, $v \in \of{E\sqb{Y} + M^\perp} \cap \of{K^M_I}^+\bs\cb{0}$ such that
$\widetilde{F}^M_{\of{Q, w}} = F^M_{\of{Y, v}}$.
\end{lemma}

{\sc Proof.} (i) Set $w = E\sqb{Y} \in \R^d_+$ since $K^+_T\of{\omega} \subseteq \R^d_+$
for every $\omega \in \Omega$, hence $L^q_d\of{K^+_T} \subseteq L^q_d\of{\R^d_+}$. Since
$v \in \of{E\sqb{Y} + M^\perp} \cap \of{K^M_I}^+\bs\cb{0}$ we have $v \in w + M^\perp$
which is equivalent to $w \in v + M^\perp$. Since $v \in \of{K^M_I}^+ = \of{K^+_I +
M^\perp} \cap M \subseteq K^+_I + M^\perp$ we may conclude $w \in K^+_I + M^\perp$. Since
$v \neq 0$ we have $w \not\in M^\perp$, hence $w \in K^+_I\bs M^\perp + M^\perp$.

Choose $Z_i = \frac{1}{w_i}Y_i$ if $w_i > 0$, and arbitrary in $\of{L^q_d}_+$ such that
$E\sqb{Z_i} = 1$ if $w_i = 0$, $i \in \cb{1, \ldots, d}$. Define $Q$ via $\frac{dQ}{dP} =
Z$. Then $Y = \diag\of{w}\frac{dQ}{dP} \in L^q_d\of{K^+_T}$ and $E\sqb{X^TY} = E\sqb{X^T
\diag\of{w}\frac{dQ}{dP}} = w^T E^Q\sqb{X}$.

By assumption $v \in w + M^\perp$, we have $v^Tu = w^Tu$ for all $u \in M$ and
\begin{equation}
\label{LemDualTransform1} F^M_{\of{Y, v}}\sqb{X} = \cb{u \in M \colon E\sqb{X^TY} \leq
v^T u} =
    \cb{u \in M \colon w^TE^Q\sqb{X} \leq w^T u} = \widetilde{F}^M_{\of{Q, w}}\sqb{X}
\end{equation}
for all $X \in L^p_d$.

(ii) Define $Y = \diag\of{w}\frac{dQ}{dP} \in L^q_d\of{K^+_T}$. Then $E\sqb{Y} = w$ and
$w^T E^Q\sqb{X} = E\sqb{X^T \diag\of{w}\frac{dQ}{dP}} = E\sqb{X^TY}$.

We claim that $w \in \of{K^M_I}^+ + M^\perp$. Indeed, on the one hand from $\of{K^M_I}^+
= \of{K^+_I + M^\perp} \cap M$ we get $\of{K^M_I}^+ + M^\perp = \of{K^+_I + M^\perp} \cap
M + M^\perp$. On the other hand, $w = w_M + w_{M^\perp}$ with $w_M \in M$, $w_{M^\perp}
\in M^\perp$, hence $w_M = w - w_{M^\perp} \in K^+_I + M^\perp$ which in turn implies $w
\in \of{K^+_I + M^\perp} \cap M + M^\perp$, and the claim is proven.

Hence, there is $v \in \of{K^M_I}^+$ such that $w \in v + M^\perp$. Since $w \not\in
M^\perp$ we have $v \neq 0$. Finally, equation \eqref{LemDualTransform1} also holds in
this case which completes the proof of the lemma. \pend

\begin{remark}
The functions $X \mapsto \widetilde{F}^M_{\of{Q, w}} \sqb{X}$ in \eqref{EqTransform} may
be seen as set-valued substitutes for the vector expectation $E^Q\sqb{X}$.
\end{remark}

\begin{remark}
If $M = \R^d$ then $v \in \of{E\sqb{Y} + M^\perp} \cap \of{K^M_I}^+\bs\cb{0}$ if and only
if $v = E\sqb{Y} \in K^+_I\bs\cb{0}$. In this case, the pair $\of{v, Y}$ is a consistent
pricing process for the one-period market model $\of{K_I, K_T}$. Compare for example
\cite{Schachermayer04} and section \ref{SubSecSH} below for definitions.
\end{remark}

\begin{remark}
The reader may note the difference between lemma \ref{LemDualTransform}, (i) above and
lemma 4.1, (i) in \cite{HamelHeyde10}. Since the cone $K_T$ is random we cannot ensure $w
= E\sqb{Y} \in K^+_I$ in general anymore. A simple counterexample is provided by example
\ref{ExWC2D} modified in the following way: Take $M = \cb{t\of{1,1}^T \colon t \in \R}$,
so that $\dim M = 1$ and $Y = E\sqb{Y} \One = \of{1,0}^T\One \in L^q_2\of{K^+_T}$. Then
$w = E\sqb{Y} \not\in K^+_I$, but $w \in K^+_I\bs M^\perp + M^\perp$. If we take the same
$w$, but use $M = \cb{t\of{0,1}^T \colon t \in \R}$ then $K^+_I\bs M^\perp + M^\perp =
\cb{x \in \R^2 \colon x_2 > 0}$ and $\of{E\sqb{Y} + M^\perp} \cap \of{K^M_I}^+\bs\cb{0} =
\emptyset$.
\end{remark}

\section{Dual representation}

\label{SecDualRep}

This section is devoted to dual representation results for market-compatible convex risk
measures. The following theorem is the main result of the paper. In order to keep its
formulation clear we shall precede it with a definition. Recall $\mathbb{G}_M = \cb{D
\subseteq \R^d \colon D = \cl\co\of{D + K^M_I}}$.

\begin{definition}
\label{DefPenalty} Define the set of dual variables
\[
\mathcal{W}^q =
    \cb{\of{Q, w} \in \mathcal{M}^P_{1,d} \times \R^d \colon
        w \in K^+_I \bs M^\perp + M^\perp, \;
        \diag\of{w}\frac{dQ}{dP} \in L^q_d\of{K^+_T}}.
\]
A function $-\alpha \colon \mathcal{W}^q \to \mathbb{G}_M$ satisfying
\\
(P0) $\bigcap_{\of{Q, w} \in \mathcal{W}^q} -\alpha\of{Q, w} \neq \emptyset$ and
$-\alpha\of{Q, w} \neq M$ for at least one $\of{Q, w} \in \mathcal{W}^q$ and
\\
(P1) $-\alpha\of{Q, w} = \cl\of{-\alpha\of{Q, w} + G\of{w}} \cap M$ for all $\of{Q, w}
\in \mathcal{W}^q$
\\
is called a penalty function.
\end{definition}

\begin{theorem}
\label{ThmDualRep} A function $R \colon L^p_d \to \mathbb{G}_M$ is a market-compatible
closed ($\sigma\of{L^\infty_d, L^1_d}$-closed if $p = \infty$) {\bf convex risk measure}
which is finite at zero if, and only if there is a penalty function $-\alpha_R$ such that
\begin{equation}
\label{ThmDualRep1}
 \forall X \in L^p_d \colon R\of{X} =
 \bigcap_{\of{Q, w} \in \mathcal{W}^q}
 \sqb{-\alpha_R\of{Q, w} + \of{E^Q\sqb{-X} + G\of{w}}\cap M}.
\end{equation}
In particular, for a closed ($\sigma\of{L^\infty_d, L^1_d}$--closed if $p = \infty$)
convex risk measure $R$ being finite at zero, \eqref{ThmDualRep1} is satisfied with
$-\alpha_R$ replaced by $-\alpha_{R, \min}$ with
\begin{equation}
\label{ThmDualRep2} -\alpha_{R, \min}\of{Q, w} = \cl\bigcup_{X' \in A_R} \of{E^Q\sqb{X'}
+ G\of{w}}\cap M.
\end{equation}
Moreover, if a penalty function $-\alpha_R$ satisfies \eqref{ThmDualRep1} then it holds
$-\alpha_R\of{Q, w} \supseteq -\alpha_{R, \min}\of{Q, w}$ for all $\of{Q, w} \in
\mathcal{W}^q$.

The function $R$ is a market-compatible closed ($\sigma\of{L^\infty_d, L^1_d}$-closed if
$p = \infty$) {\bf coherent risk measure} which is finite at zero if, and only if there
is a nonempty set $\mathcal{W}^q_R \subseteq \mathcal{W}^q$ such that
\begin{equation}
\label{ThmDualRepCoh1}
 \forall X \in L^p_d \colon R\of{X} =
 \bigcap_{\of{Q, w} \in \mathcal{W}^q_R} \of{E^Q\sqb{-X} + G\of{w}}\cap M.
\end{equation}
In particular, \eqref{ThmDualRepCoh1} is satisfied with $\mathcal{W}^q_R$ replaced by
$\mathcal{W}^q_{R, \max}$ with
\begin{equation}
\label{ThmDualRepCoh2} \mathcal{W}^q_{R, \max} =
 \cb{\of{Q, w} \in \mathcal{M}^P_{1,d} \times \R^d \colon w \in K^+_I \bs M^\perp
 + M^\perp, \; \diag\of{w}\frac{dQ}{dP} \in A^+_R}.
\end{equation}
Moreover, if $\mathcal{W}^q_R$ satisfies \eqref{ThmDualRepCoh1} then $\mathcal{W}^q_R
\subseteq \mathcal{W}^q_{R, \max}$.
\end{theorem}

{\sc Proof.} See appendix. \pend

\medskip Theorem \ref{ThmDualRep} together with remark \ref{RemPreAcceptSupport}
produces the following possibility for generating market-compatible convex risk measures.
Starting with a nonempty convex set $\widehat{A} \subseteq L^p_d$ one has to ensure that
$A = \cl\of{\widehat{A} + L^p_d\of{K_T} + K^M_I\One}$ satisfies (A1b) (definition
\ref{DefAccSet}, $A$ satisfies (A1a) if $\widehat{A}$ does). Then \eqref{ThmDualRep1}
with $-\alpha_R$ replaced by
\begin{equation}
\label{EqDualRepPreAcc} -\alpha_{R_A}\of{Q, w} = \cl\bigcup_{X \in A} \of{E^Q\sqb{X} +
G\of{w}}\cap M
\end{equation}
produces a closed convex market-compatible risk measure which is finite at zero. This
will be useful, for example, when we consider $\widehat{A} = L^p_d\of{K_T}$ where the
robust no-arbitrage assumption for the one-period market model $\of{K_I, K_Y}$ ensures
that already $A = L^p_d\of{K_T} + K^M_I\One$ is a closed convex acceptance set. See
section \ref{SecWCRF} below.

\begin{remark}
Consider the special case $m = d$, $K_I = K^M_I = \R^d_+ \equiv K_T$, i.e. the totally
illiquid market. Theorem \ref{ThmDualRep} produces a dual representation of a closed
convex regulator risk measure with
\[
\mathcal{W}^q =
    \cb{\of{Q, w} \in \mathcal{M}^P_{1,d} \times \R^d \colon
        w \in \R^d_+\bs\cb{0}, \;
        \diag\of{w}\frac{dQ}{dP} \in L^q_d\of{\R^d_+}}.
\]
Thus, the set of dual variables is larger than for any other market in the case $d = m$
since $R^d_+ \subseteq K_I$, $R^d_+ \subseteq K_T$ a.s., hence $K^+_I, K^+_T \subseteq
\R^d_+$ a.s. Such a regulator risk measure "ignores the market" in the sense that the
results of trades, possible according to the "real" market model, might not be included
in the set of risk compensating initial endowments for a given position $X$.
\end{remark}

\section{Examples}

\label{SecExamples}

\subsection{A remark about scalarization}

In \cite{HamelHeyde10}, we introduced a scalarization procedure for functions $R \colon
L^p_d \to \mathcal{P}\of{M}$ by defining extended real-valued functions $\vp_{R, v}
\colon L^p_d \to \R\cup\cb{\pm\infty}$ given by
\[
\vp_{R, v}\of{X} = \inf_{u \in R\of{X}} v^Tu
\]
for $v \in K^+_I$. The following simple observation will be useful for the example in
section \ref{SubSecFrictless}.

\begin{proposition}
\label{PropScalarPreRisk} Let $\widehat{A} \subseteq L^p_d$ be an acceptance set and $A =
\cl_M\of{\widehat{A} + K^M_I\One}$. Then
\[
\forall X \in L^p_d \colon \vp_{R_A, v}\of{X} = \vp_{R_{\widehat{A}}, v}\of{X}
\]
whenever $v \in K^+_I$.
\end{proposition}

{\sc Proof.} Since $R_A\of{X} \supseteq R_{\widehat{A}}\of{X}$ we have $\leq$. To show
the converse, take $u \in M$ such that $X + u\One \in \widehat{A} + K^M_I\One$. That is,
there is $k \in K^M_I$ such that $X + \of{u-k}\One \in \widehat{A}$ which is $u-k \in
R_{\widehat{A}}\of{X}$. Hence
\[
\vp_{R_{\widehat{A}}, v}\of{X} \leq v^T\of{u-k} \leq v^Tu
\]
since $k \in K^M_I \subseteq K_I$ and $v \in K^+_I$. A continuity argument yields that
this is also true for $u \in M$ with $X + u\One \in A = \cl_M\of{\widehat{A} +
K^M_I\One}$. Hence $\vp_{R_A, v}\of{X} \geq  \vp_{R_{\widehat{A}}, v}\of{X}$ as desired.
\pend

\medskip We note that the optimization problem to determine the value $\vp_{R_A, v}\of{X}$
may have many more solutions than the problem to determine $\vp_{R_{\widehat{A}},
v}\of{X}$ despite the fact that the optimal values for both problems are the same. In
fact, this is usually the case as the following example illustrates.

\subsection{The case of one leading currency and several eligible assets}
\label{SubSecFrictless}

In their 2009 paper \cite{ArtDelKochMed09} the authors introduce the following
generalization of scalar monetary measures of risk. Assume we are given a "leading"
(domestic) currency (eligible asset number 1) and a set of $d-1$ further eligible assets.
One unit of asset $i$ has the deterministic price $\pi_i > 0$ at initial time (with
$\pi_1 = 1$ and the random price $S_i \colon \Omega \to \R_+\bs\cb{0}$ at terminal time,
$i \in \cb{1, \ldots, d}$ (with $S_1 \equiv 1$ while assuming that money in the leading
currency is invested in a risk free manner and everything is discounted). An eligible
portfolio $x = \of{x_1, \ldots, x_d} \in \R^d$ (in "physical" units) has the price
$\sum_{i=1}^d \pi_ix_i$ at initial time and $S = \sum_{i=1}^d S_ix_i$ at terminal time.
Let us further assume that we are given a non-empty set $A_1 \subseteq L^0_1$ of
acceptable random payoffs in the leading currency (we do not call it an acceptance set
since we do not require further properties yet). The function
\[
\varrho\of{Z} = \inf\cb{s \in \R \colon Z + \sum_{i=1}^d S_ix_i \in A_1, \; x \in \R^d,
\; \sum_{i=1}^d \pi_ix_i = s}, \; Z \in L^0_1,
\]
corresponds to the risk measure of definition 2 in \cite{ArtDelKochMed09} if the set
$\mathcal{S}$ therein has the $d$ members described above.

We reformulate the above model in terms of our previous constructions. We have $m=d$, and
the solvency cone at initial time is the polyhedral convex cone $K_I$ spanned by the
$d-1$ vectors
\[
\of{-1, 0, \ldots, 0, \frac{1}{\pi_i}, 0, \ldots, 0}, \; i = 2, \ldots, d,
\]
their negatives and the vector $\of{1, \ldots, 1} \in \R^d$. The values of the solvency
cone mapping $K_T$ at terminal time are spanned by
\[
\of{-1, 0, \ldots, 0, \frac{1}{S_i}, 0, \ldots, 0}, \; i = 2, \ldots, d,
\]
their negatives and the vector $\of{\One, \ldots, \One} \in L^p_d$. Both cones are closed
half spaces (the second one random) with normal vectors
\[
\of{1, \pi_2, \ldots, \pi_d} \; \mbox{and} \;
    \of{\One, S_2, \ldots, S_d},
\]
respectively. That is, we are moving in a frictionless market. Define a set $\widehat{A}
\subseteq L^0_d$ of acceptable positions by
\[
X = \of{X_1, \ldots, X_d} \in \widehat{A} \quad \Longleftrightarrow \quad
    \sum_{i=1}^d S_iX_i \in A_1.
\]
It should be clear that $\widehat{A}$ is convex (a cone, $K_T$-compatible) if and only if
$A_1$ is convex (a cone, $\of{L^p_1}_+$-compatible). However, $\widehat{A}$ is not
market-compatible in general even if $A_1$ is an acceptance set in $L^p_1$. For a random
payoff $X_1 \colon \Omega \to \R$ and $u \in \R^d$ we have
\[
\of{X_1, 0, \ldots, 0} + u\One \in \widehat{A}
    \quad \Longleftrightarrow \quad X_1 + \sum_{i=1}^d S_iu_i \in A_1.
\]
The collection
\[
\cb{u \in \R^d \colon \of{X_1, 0, \ldots, 0} + u\One \in \widehat{A}} =
    \cb{u \in \R^d \colon  X_1 + \sum_{i=1}^d S_iu_i \in A_1}
\]
is the value of a set-valued risk measure $R_{\widehat{A}}$ at $\of{X_1, 0, \ldots, 0}$,
and $R_{\widehat{A}}$ is $K_T$-compatible. Moreover,
\[
\varrho\of{X_1} = \inf\cb{\sum_{i=1}^d \pi_iu_i \colon u \in R_{\widehat{A}}\of{\of{X_1,
0, \ldots, 0}}}.
\]
Thus, $\varrho$ is nothing else than the scalarization of $R_{\widehat{A}}$ (see the
previous subsection) with $v = \pi \in K^+_I$. By proposition \ref{PropScalarPreRisk}
$\varrho$ is also a scalarization of the risk measure $R_A$ with $A =
\cl_M\of{\widehat{A} + K_I\One}$. This shows that the risk measures defined in
\cite{ArtDelKochMed09} fit perfectly into the framework of the present paper. In fact,
the situation considered in \cite{ArtDelKochMed09} is a very special case of our model:
The idea of investing in eligible assets is covered by the set-valued approach presented
here. We illustrate this below with a toy example taken from \cite{ArtDelKochMed09}.

Moreover, the approach of \cite{ArtDelKochMed09} can easily be generalized to situations
(a) where proportional frictions are present together with (b) when there are future
payoffs not only in the "leading" (domestic) currency and (c) when some assets are
illiquid in the sense that they can not be exchanged into certain others (in this case,
some faces of the solvency cones $K_I$ and/or $K_T$ coincide with certain faces of
$\R^d_+$, this would mean $\pi_i = +\infty$ or $S_i\of{\omega} = +\infty$ for some $i$ in
the framework of \cite{ArtDelKochMed09}).

Finally, we should note that the question of a dual representation of the risk measures
defined in \cite{ArtDelKochMed09} can easily be solved using the results of section
\ref{SecDualRep}. This problem was not addressed in \cite{ArtDelKochMed09}.

The following example is the first in the appendix of \cite{ArtDelKochMed09}. Take a
probability space with $\Omega = \cb{\omega_1, \omega_2, \omega_3}$. Consider the case $d
= m = 2$, the (six dimensional) payoff space $L^1_2\of{\Omega, 2^\Omega, P}$ and the
following market model: $K_I = \cb{x \in \R^2 \colon x_1 + x_2 \geq 0}$ (one-to-one
exchange between the two assets without transaction costs) and
\[
K_T\of{\omega_1} = \cb{x \in \R^2 \colon x_1 + 2x_2 \geq 0}, \;
    K_T\of{\omega_2} = K_I, K_T\of{\omega_3} = \cb{x \in \R^2 \colon 2x_1 + x_2 \geq 0}.
\]
The payoff is given by $X = \of{X_1, 0}$ with $X_1\of{\omega_1} = -16$, $X_1\of{\omega_2}
= 1$ and $X_1\of{\omega_3} = -7$. We have
\[
\cb{u \in \R^2 \colon X + u\One \in L^1_2\of{K_T}} =
    \cb{u \in \R^2 \colon u_1 + 2u_2 \geq 16, \;  2u_1 + u_2 \geq 14}.
\]
The unique minimal (in the sense of vector optimization, see \cite{Jahn04}, often called
efficient) point with respect to the reflexive, transitive relation generated by the cone
$K_I$ (actually, a half space) is indeed the point $\of{4,6}$ computed in
\cite{ArtDelKochMed09}. Obviously, transaction costs can be introduced in the example in
such a way that the main features remain unchanged ("small" transaction costs) or such
that the whole boundary of the above set becomes "efficient".

Note that $L^1_2\of{K_T}$ is not a market-compatible acceptance set, but the set
$L^1_2\of{K_T} + K_I\One$ is. Considering the set
\begin{multline*}
\cb{u \in \R^2 \colon \exists k \in K_I \colon X + \of{u-k}\One \in L^1_2\of{K_T}} \\ =
    \cb{u \in \R^2 \colon u_1 - k_1 + 2u_2 - 2k_2 \geq 16, \;  2u_1 - 2k_1 + u_2 - k_1 \geq 14, \;
    k_1 + k_2 \geq 0}
\end{multline*}
as the feasible set for a linear program with objective function $u_1 + u_2$ (to be
minimized) we obtain from proposition \ref{PropScalarPreRisk} that the value of this
program is again 10. The set of solutions, however, is
\[
\cb{u \in \R^2 \colon u_1 = t + 4, \; u_2 = -t + 6, \; t \in \R}.
\]
This set consists of all initial positions which can be exchanged at initial time into
$\of{4,6}$. One should expect that a mathematical model admits to detect all such
positions since they are trivially "as good as" the position $\of{4,6}$. Thus, even if
there are no transaction costs present, but random exchange rates it makes sense to use
the set-valued approach -- because the latter yields more solutions among which an agent
may choose.

\subsection{Value at risk}

We will give a general definition of value at risk which extends the one in
\cite{HamelRudloff08} and \cite{HamelHeyde10} to the case of a random solvency cone at
terminal time.

\begin{definition}
\label{DefVaR} Let $\alpha \in \sqb{0,1}$ and a measurable set-valued mapping $D_\alpha
\colon \Omega \to \mathcal{P}\of{\R^d}$ be given which satisfy
\\
(D0) $D_\alpha\of{\omega}$ is a non-empty closed convex set for each $\omega \in \Omega$,
\\
(D1a) $\R^d_+ \subseteq D_\alpha\of{\omega}$ $P$-a.s.,
\\
(D1b) $-\Int \R^d_+ \cap D_\alpha\of{\omega} = \emptyset$ $P$-a.s.,
\\
(D2) $D_\alpha\of{\omega} + \R^d_+ \subseteq D_\alpha\of{\omega}$ $P$-a.s.
\\
The value at risk $V@R_{\alpha}$ with respect to $D_\alpha$ is defined by
\[
V@R_{\alpha}\of{X} =
    \cb{u\in M \colon X + u\One \in \widehat{A}_{D_\alpha}}, \quad X \in L^p_d,
\]
where
\[
\widehat{A}_{D_\alpha} =
    \cb{X \in L^p_d \colon P\of{\cb{\omega \in \Omega \colon
        X\of{\omega} \in D_\alpha\of{\omega}}} \geq 1-\alpha}.
\]
\end{definition}

It is easily seen that $\widehat{A}_{D_\alpha}$ is an acceptance set in $L^p_d$, $0 \leq
p \leq \infty$. If the requirement (D2) is replaced by
\\
(D2') $D_\alpha\of{\omega} + K_T\of{\omega} \subseteq D_\alpha\of{\omega}$ $P$-a.s.,
\\
then
\[
A_{D_\alpha} = \widehat{A}_{D_\alpha} + K^M_I\One
\]
is a market-compatible acceptance set, hence the corresponding risk measure is a
market-compatible value at risk. Note that condition (A1b) is not satisfied in general as
the example with $m = 1$, $d =2$, $D_\alpha\of{\omega} = \cb{x \in \R^2 \colon x_2 \geq
0}$, $M = \R \times \cb{0}$ shows. An assumption like $M \cap \of{\R^d \bs
D_\alpha\of{\omega}} \neq \emptyset$ $P$-a.s. avoids this situation.

The definition of value at risk in the case of random terminal prices is more cumbersome
than the versions in \cite{HamelRudloff08}. Note also that the sets $D_\alpha\of{\omega}$
above correspond to the complement of the set $D$ used in \cite{HamelRudloff08}.

A choice for the mapping $D_\alpha$ which seems natural would be $D_\alpha\of{\omega} =
K_T\of{\omega}$. This $D_\alpha$ satisfies (D0), (D1a), (D1b) and (D2').

\subsection{The worst case risk measure}

\label{SecWCRF}

The worst case risk measure should be generated by the smallest possible acceptance set
which includes $0 \in L^p_d$. This is $\of{L^p_d}_+$ which certainly is an acceptance set
in the sense of definition \ref{DefAccSet}. The augmented, market-compatible acceptance
set is
\[
A^p_{WC} = \cl_M\of{L^p_d\of{K_T} + K^M_I\One},
\]
which generates the risk measure
\[
WC^p\of{X} = \cb{u \in M \colon X + u\One \in A^p_{WC}} =
    \cl\cb{u \in M \colon X + u\One \in L^p_d\of{K_T} + K^M_I\One}.
\]
The set $A^p_{WC}$ obviously satisfies $M\One \cap A^p_{WC} \neq \emptyset$ (see (A1a)),
and is market-compatible (use lemma \ref{LemDirClosure1}).

A sufficient condition for $M\One \cap \of{L^p_d \bs A} \neq \emptyset$ (see  (A1b)), is
no arbitrage in the market (see \cite[Definition 1.6.]{Schachermayer04}, \cite[p.
161]{PennanenPenner08R}), that is for our case of a one-period market
\[
\of{L^p_d\of{K_T} + K_I\One} \cap -L^p_d\of{\R^d_+} = \cb{0}
\]
since this implies $\of{u - k}\One \not\in L^p_d\of{K_T}$ for every $u \in
-\R^d_+\bs\cb{0}$ and every $k \in K_I$. Hence (A1b) follows from
$\of{-\R^d_+\bs\cb{0}}\cap M \neq \emptyset$.

Moreover, it is known that the robust no arbitrage condition (see \cite[Definition
1.9]{Schachermayer04}, \cite[Definition 3.1]{PennanenPenner08R}) implies that the set
$L^p_d\of{K_T} + K_I\One$ is closed. In this case one could drop the closure in
$A^p_{WC}$ below. In general, for $p \geq 1$ the set
\[
A^p_{WC} = \cl\of{L^p_d\of{K_T} + K^M_I\One}
\]
is a nontrivial closed convex cone in $L^p_d$ defining a closed coherent risk measure.
The dual cone of this cone in $L^p_d$ is the set
\[
\of{A^p_{WC}}^+ = \cb{Y \in L^q_d\of{K^+_T} \colon E\sqb{Y} \in K^+_I + M^\perp}
\]
as one easily verifies.

On the other hand, according to theorem \ref{ThmDualRep1}, equations
\eqref{ThmDualRepCoh1} and \eqref{ThmDualRepCoh2} we have for the corresponding risk
measure
\[
\forall X \in L^p_d \colon WC^p\of{X} =
 \bigcap_{\of{Q, w} \in \mathcal{W}^q_{max}} \widetilde{F}^M_{\of{Q, w}}\sqb{-X}
\]
with
\[
\mathcal{W}^q_{max} =
 \cb{\of{Q, w} \in \mathcal{M}^P_{1,d} \times \R^d \colon w \in \of{K^+_I \bs M^\perp
 + M^\perp}, \; \diag\of{w}\frac{dQ}{dP} \in \of{A^p_{WC}}^+}.
\]
Actually, we have $\mathcal{W}^q_{max} = \mathcal{W}^q$ (see definition
\ref{DefPenalty}). Indeed, take $\of{Q, w} \in \mathcal{W}^q$, i.e. $w \in \of{K^+_I \bs
M^\perp} + M^\perp$ and $\diag\of{w}\frac{dQ}{dP} \in L^q_d\of{K^+_T}$. Then $Y =
\diag\of{w}\frac{dQ}{dP} \in \of{A^p_{WC}}^+$. Indeed, taking $X \in L^p_d\of{K_T}$ and
$k \in K^M_I$ we obtain
\[
E\sqb{Y^T\of{X + k\One}} = E\sqb{X^T\diag\of{w}\frac{dQ}{dP}} + k^Tw \geq 0
\]
because of the properties of $Y$ and $w$. Thus, $\mathcal{W}^q \subseteq
\mathcal{W}^q_{max} $. The converse inclusion is trivial since $L^p_d\of{K_T} \subseteq
A^p_{WC}$.

In the case $M = \R^d$ this risk measure provides a link to superhedging theorems for
conical markets models as shown in subsection \ref{SubSecSH} below.

\subsection{Average value at risk}

We will give a dual definition of Average Value at Risk which slightly differs from the
definition given in \cite{HamelHeyde10}. For $\lambda\in(0,1]^d$ we define the vector
\[
\mu\of{\lambda} = \of{\frac{1}{\lambda_1}, \frac{1}{\lambda_2}, \ldots,
\frac{1}{\lambda_d}}^T
\]
and the set
\[
\mathcal{W}^\infty_\lambda = \cb{\of{Q, w} \in \mathcal{W}^\infty \colon
\diag\of{w}\of{\mu\of{\lambda}\One - \frac{dQ}{dP}} \in L^\infty_d\of{K^+_T}}.
\]
In contrast to the constant cone case $K_I \equiv K_T$, it may happen that the set
$\mathcal{W}^\infty_\lambda$ is empty. If $\mathcal{W}^\infty_{\lambda} \neq \emptyset$
then
\[
AV@R_\lambda\of{X} = \bigcap_{\of{Q,w} \in \mathcal{W}^\infty_\lambda}
        \of{E^Q\sqb{-X}+G\of{w}}\cap M
\]
defines a market-compatible coherent risk measure according to theorem \ref{ThmDualRep}.

\subsection{Superhedging prices as a coherent risk measure}

\label{SubSecSH}

In this section, we show that the set of superhedging prices for a multivariate claim in
a market with proportional transaction costs can be understood as value of a set-valued
coherent risk measure. This is in complete analogy to the frictionless case, compare for
example \cite[sections 1.3, 4.8]{FoellmerSchied04}.

In our notation, we mainly follow \cite{Schachermayer04}. We consider a financial market
with $d$ securities which can be traded over finite discrete time $t=0, \ldots, T$. The
information evolves according to a filtration $\of{\mathcal F_t}_{t=0}^T$ on a
probability space $\of{\Omega, \mathcal F, P}$ satisfying the usual conditions. In
particular, it is assumed that $\mathcal F_0$ is the trivial $\sigma$-algebra.

The market model is given by the solvency cone process $\of{K_t\of{\omega}}_{t= 0}^T$
where $K_t\of{\omega}$ is a closed convex cone with $\R^d_+ \subseteq K_t\of{\omega} \neq
\R^d$ for all $t=0, \ldots, T$ and all $\omega\in\Omega$. The solvency cone $K_t$ is the
collection of positions transferrable into nonnegative positions by taking into account
the transaction costs at time $t$.

In contrast to the rest of the paper, where only the two time points $t=0$ and $t=T$ were
important, we are now also interested in trading at intermediate time points. Thus, not
only $K_I = K_0$ and $K_T$ play a role, but also the (random) solvency cones $K_t$ at
intermediate time points will be of interest.

A portfolio vector is a random variable $V_t \colon \Omega \to \R^d$. The values
$V_t\of{\omega}$ of portfolio vectors are given in physical units, i.e.,
$V_{t,i}\of{\omega}$ is the number of units of the $i$th asset for $i = 1, \ldots, d$ in
the portfolio at time $t$.

An $\R^d$-valued adapted process $\of{V_t}_{t=0}^T$ is called a self-financing portfolio
process for the market given by $\of{K_t}_{t=0}^T$ iff
\[
\forall t=0,\dots, T \colon V_t - V_{t-1} \in -K_t \quad P-\mbox{a.s.}
\]
with the convention $V_{-1}=0$.

For $t=0, \ldots, T$, we denote by $A_t \subseteq L^0_d\of{\Omega, \mathcal{F}_t, P}$ the
set of random vectors $V_t \colon \Omega \to \R^d$, each being the value of a
self-financing portfolio process at time $t$, i.e. $A_t$ is the set of super-hedgeable
claims starting from initial endowment $0\in\R^d$. As it easily follows from the
definition of self-financing portfolio processes, $A_t$ is a convex cone.

The fundamental theorem of asset pricing states that the robust no arbitrage condition is
satisfied for the market model $\of{K_t}_{t=0}^T$ if and only if there exists a strictly
consistent pricing process $\of{Z_t}_{t = 0}^T$  (see \cite{Schachermayer04}, theorem~1.7
for the case of a polyhedral market model and \cite{PennanenPenner08R}, theorem~18 for
the general case). The definitions are as follows.

The market given by $\of{K_t}_{t=0}^T$ is said to satisfy the robust no--arbitrage
property (NA$^\text{r}$) iff there exists a market process $\of{\widetilde{K}_t}_{t=0}^T$
satisfying
\begin{equation}
\label{EqRNA}
 K_t \subseteq \widetilde{K}_t \quad \mbox{and} \quad
    K_t \bs -K_t \subseteq \Int \widetilde{K}_t\quad P-\mbox{a.s.}
\end{equation}
for all $t=0,\ldots, T$ such that
\[
\widetilde{A}_T \cap L^0_d\of{\R^d_+} = \cb{0},
\]
where $\widetilde{A}_T$ is generated by the self-financing portfolio processes with
\[
\forall t=0, \ldots, T \colon\quad V_t - V_{t-1} \in -\widetilde{K}_t \quad P-\mbox{a.s.}
\]
An $\R^d_+$-valued adapted process $Z = \of{Z_t}_{t = 0}^T$ is called a (strictly)
consistent pricing process for the market model $\of{K_t}_{t=0}^T$ if $Z$ is a martingale
under $P$ and
\[
\forall t=0,\dots, T \colon\quad
    Z_t \in K^+_t\bs\cb{0} \quad \of{\in \ri K^+_t} \;
P-\mbox{a.s.}
\]
The superhedging price of a claim $X \in L^0_d$ can be characterized as follows.

\begin{theorem}[\cite{Schachermayer04}, theorem~4.1, \cite{PennanenPenner08R}, corollary~15]
\label{ThmSH} Assume that the market process $\of{K_t}_{t=0}^T$ satisfies the robust
no-arbitrage condition (NA$^\text{r}$). Then, the following conditions are equivalent for
$X \in L^0_d$ and $u^0 \in \R^d$:

(i) $X - u^0 \in A_T$, i.e. there is a self-financing portfolio process
$\of{V_t}_{t=0}^T$ such that
\[
u^0 + V_T = X \quad P-\mbox{a.s.}
\]

(ii) For every consistent (or, equivalently, strictly consistent) pricing process $Z =
\of{Z_t}_{t=0}^T$ such that the negative part $\of{X^TZ_T}^-$ is integrable it holds
\[
E\sqb{X^TZ_T} \leq \of{u^0}^TZ_0.
\]
\end{theorem}

{\sc Proof.} This follows from theorem 18 in \cite{PennanenPenner08R} in the same way as
Theorem 4.1 follows from theorem 1.7 in \cite{Schachermayer04} if one observes that lemma
2.5 in \cite{Schachermayer04} can be replaced by condition \eqref{EqRNA}. \pend

\medskip An element $u^0 \in \R^d$ satisfying the conditions of theorem \ref{ThmSH} is called
a superhedging price of $X$. Now, we shall show that the map assigning to $X$ the
collection of its superhedging prices defines a closed coherent risk measure on $L^0_d$.

\begin{corollary}
\label{CorRiskViaSH} An element $u^0 \in \R^d$ is a superhedging price for the claim $X
\in L^0_d$ if and only if $u^0\in R_{-A_T}\of{-X}$ with
\begin{equation}
\label{SHprimal L^0}
   R_{-A_T}\of{-X} := \cb{u\in \R^d \colon -X + u\One \in -A_T}.
\end{equation}
If the market process $\of{K_t}_{t=0}^T$ satisfies the robust no-arbitrage condition
(NA$^\text{r}$), then $R_{-A_T}$ is a closed coherent market-compatible risk measure on
$L^0_d$ and has the following representation
\begin{equation}
\label{SHdual}
 R_{-A_T}\of{-X} = \bigcap_{\cb{\of{Q, w} \in \mathcal{W}^1_{\cb{0, \ldots, T}} \colon
    E^Q\sqb{X^-} < \infty}} \of{E^Q\sqb{X} + G\of{w}},
\end{equation}
where
\begin{align*}
\mathcal{W}^1_{\cb{0, \ldots, T}} = & \big\{\of{Q,w} \in \mathcal{M}^P_{1,d}\times
\R^d\bs\cb{0} \colon
\\
& E\sqb{\diag\of{w}\frac{dQ}{dP} \Big| \mathcal{F}_t} \in L^1_d\of{\Omega, \mathcal{F}_t,
P; K_t^+}\mbox{ for all } t = 0, \ldots, T \big\}.
\end{align*}
\end{corollary}

{\sc Proof.} Condition (i) of theorem~\ref{ThmSH} leads to the first assertion. The set
$-A_T$ is an acceptance set in the sense of definition \ref{DefAccSet}. Indeed, the first
condition is trivial, the second one follows from the the robust no-arbitrage condition
(NA$^\text{r}$) and the third one holds since by definition, $-A_T = K_0\One +
L^0_d\of{\Omega, \mathcal{F}_1, P;K_1} + \ldots + L^0_d\of{\Omega, \mathcal{F}_T, P;K_T}$
and $K_t\of{\omega}$ is a convex cone with $\R^d_+\subseteq K_t\of{\omega}$ for $t=0,
\dots, T$ and all $\omega \in \Omega$. This last condition also implies that $-A_T$ is
$K_t$-compatible (see definition \ref{DefMarketComp}) for all $t=0, \dots, T$ and thus
market-compatible. Furthermore, $-A_T \subseteq L^0_d$ is a convex cone and closed in
$L^0_d$ (see \cite{Schachermayer04}, theorem~2.1 for the polyhedral case and
\cite{PennanenPenner08R}, theorem~7 for the general case), thus also directionally
closed. Via the primal representation \eqref{EqRA} with $M = \R^d$, $-A_T$ defines a risk
measure $R_{-A_T}$ on $L^0_d$, which is closed coherent and market-compatible according
to proposition \ref{PropMarketComp}, \ref{PropConvexity} and \ref{PropClosedness}.

By theorem~\ref{ThmSH}~(ii) the set $R_{-A_T}\of{-X}$ of superhedging prices of $X$ can
also be written in the following form
\begin{equation}
    \label{SH_CPP}
    R_{-A_T}\of{-X} = \bigcap_{Z \in CPP} \cb{u \in \R^d \colon E\sqb{X^TZ_T} \leq Z^T_0u},
\end{equation}
where $CPP$ is the set of consistent pricing processes $Z = \of{Z_t}_{t=0}^T$ such that
the negative part $\of{X^TZ_T}^-$ is integrable. The idea of lemma~\ref{LemDualTransform}
can be applied to consistent pricing processes in order to get the representation
\eqref{SHdual}. Indeed, starting with a consistent pricing process $Z$, one obtains a
pair $\of{Q,w} \in \mathcal{W}^1_{\cb{0, \ldots, T}}$ by defining $w := E\sqb{Z_T}= Z_0
\in K^+_0\bs\cb{0}$ and
\[
\frac{dQ_i}{dP} := \frac{1}{w_i}\of{Z_T}_i \quad \mbox{if} \; w_i > 0,
\]
for $i=1,\dots, d$ and choosing $\frac{dQ_i}{dP}$ as a density of a probability measure
in $L^1_d(\R^d_+)$ if $w_i = 0$. Conversely, a pair $(Q,w)\in\mathcal{W}^1_{\cb{0,
\ldots, T}}$ yields a consistent pricing process $Z$ by setting $Z_T =
\diag\of{w}\frac{dQ}{dP}$ and $Z_t = E\sqb{Z_T | \mathcal{F}_t}$ for $t = 0,\ldots, T$.
Together with the representation \eqref{SH_CPP} this leads to \eqref{SHdual}. \pend

\medskip Note that for $X\in L^0_d$, the set of dual elements in \eqref{SHdual} depends on the
particular argument $X$. This is no longer the case for $X\in L^p_d$, $1\leq p\leq
\infty$ as we will see below in equation \eqref{SHdualp}.

\begin{corollary}
Assume (NA$^\text{r}$). For $X\in L^p_d$, $1\leq p\leq \infty$, the superhedging prices
of $X$ form the value of a closed coherent market-compatible risk measure on $L^p_d$ with
primal representation
\begin{equation*}
    R_p\of{-X} := \cb{u \in \R^d \colon -X + u\One \in -A^p_T},
\end{equation*}
where $A^p_T := A_T \cap L^p_d$, and dual representation
\begin{equation}
\label{SHdualp} R_p\of{-X} = \bigcap_{\of{Q, w} \in \mathcal{W}^q_{\cb{0, \ldots, T}}}
\of{E^Q\sqb{X} + G\of{w}},
\end{equation}
where
\begin{align}
\nonumber \mathcal{W}^q_{\cb{0, \ldots, T}} = &\Big\{\of{Q, w} \in
\mathcal{M}^P_{1,d}\times \R^d\bs\cb{0} :
\\
\label{dyn_coupling} & E\sqb{\diag\of{w}\frac{dQ}{dP} \Big| \mathcal{F}_t} \in
L^q_d(\Omega, \mathcal{F}_t, P;K_t^+)\mbox{ for all } t = 0, \ldots, T \Big\}.
\end{align}
\end{corollary}

{\sc Proof.} This follows since $-A^p_T := -A_T \cap L^p_d$ is a closed convex cone and a
market-compatible acceptance set for $1\leq p\leq \infty$ (if $p = \infty$ the
$\sigma(L^\infty_d, L^1_d)$ closedness follows with help of lemma A.64 in
\cite{FoellmerSchied04}). The dual representation follows from theorem~\ref{ThmDualRep}
in connection with corollary \ref{CorRiskViaSH}. \pend

\medskip In analogy to the frictionless case, equations~\eqref{SHdual} and \eqref{SHdualp} show that
the superhedging prices for a multivariate claim $X$ form indeed the value of a
set-valued coherent risk measure at $-X$.

\begin{remark}
In contrast to theorem~\ref{ThmDualRep}, a dynamic coupling condition appears in
\eqref{dyn_coupling}, which is due to the fact, that also the intermediate time points
$t=0, \ldots, T$ are important and enter the risk measure through its acceptance set. The
dynamic coupling condition in \eqref{dyn_coupling} ensures that the pricing process
$\of{Z_t}_{t=0}^T$ corresponding to $(Q,w)$ is consistent with the market
$\of{K_t}_{t=0}^T$ at every time point $t=0, \ldots, T$. Note that $M^\bot=\{0\}$, and
the dynamic coupling condition in \eqref{dyn_coupling} for $t=0$ implies that $w \in
K^+_0\bs\cb{0}$ as well as $\diag\of{w}\frac{dQ}{dP} \in L^q_d\of{K^+_T}$ for $t = T$.
\end{remark}

\begin{remark} \label{EMM}
The vector probability measures $Q$ with $\of{Q, w} \in \mathcal{W}^1_{\cb{0, \ldots,
T}}$ can be seen as equivalent martingale measures. Indeed, the component $Q_i$, $i = 1,
\ldots, d$, is an equivalent martingale measure in the sense of \cite{DelSch06} if asset
$i$ is chosen as num\'{e}raire.
Instead of choosing a particular num\'{e}raire in advance, we work with all possible
num\'{e}raires, i.e. with all $Q_1, \ldots, Q_d$ representing one (scalar) equivalent
martingale measure, at the same time.
\end{remark}

\section{Appendix}

\subsection{Directionally closed sets in $L^p_d$}

As before, we use the symbol $u^k \stackrel{M}{\longrightarrow} 0$ in order to denote a
sequence $\cb{u^k}_{k \in \N} \subset M$ with $\lim_{k \to \infty} u^k = 0$. Recall (see
definition \ref{DefRadCloSet}) that the set
\[ \cl_M A = \cb{X \in L^p_d \colon
\exists u^k \stackrel{M}{\longrightarrow} 0 \colon
    \forall k \in \N \colon X + u^k\One \in A}
\]
is called the directional closure of $A$ in $M$. The following immediate results show
that this definition makes sense.

\begin{lemma}
\label{LemDirClosure1} If $A \subseteq L^p_d$, then $A \subseteq \cl_M A$, and $\cl_M A$
is directionally closed in $M$. Moreover, if $A, B \subseteq L^p_d$, then $\cl_M A +
\cl_M B \subseteq \cl_M\of{A + B}$.
\end{lemma}

{\sc Proof.} By definition, $A \subseteq \cl_M A$. Next, we shall show that $\cl_M A$ is
directionally closed. Take $X \in L^p_d$ and $u^n \stackrel{M}{\longrightarrow} 0$ such
that
\[
\forall n \in \N \colon X + u^n\One \in \cl_M A.
\]
Then, for each $n \in \N$ there is $v^{n,k} \stackrel{M}{\longrightarrow} 0$ (for $k \to
\infty$) such that
\[
\forall k \in \N \colon X + \of{u^n + v^{n,k}} \One \in A.
\]
Define a new sequence $\cb{w^n}_{n\in \N} \subseteq M$ by setting $w^1 = u^1 + v^{1,1}$
and $w^n = u^n + v^{n, k_n}$ for $n = 2,3, \ldots$ such  that $\abs{v^{n, k_n}} \leq
\frac{1}{2}\abs{v^{n-1, k_{n-1}}}$. This is possible since $v^{n,k} \to 0$ as $k \to
\infty$ for all $n \in \N$. Thus, $w^n = u^n + v^{n, k_n}  \stackrel{M}{\longrightarrow}
0$ and
\[
\forall n \in \N \colon X + w^n\One \in A.
\]
Hence $X \in \cl_M A$ by definition of the directional closure. For the last claim, take
$X \in \cl_M A + \cl_M B$. Then there are $X^A \in \cl_M A$ and $X^B \in \cl_M B$ such
that $X = X^A + X^B$. By definition of the directional closure there are sequences $u^k
\stackrel{M}{\longrightarrow} 0$ and $v^k \stackrel{M}{\longrightarrow} 0$ such that
\[
\forall k \in \N \colon X^A + u^k\One \in A, \; X^B + v^k\One \in B.
\]
Hence $X^A + X^B +\of{u^k + v^k}\One \in A + B$ which implies $X^A + X^B \in \cl_M\of{A +
B}$ since $\of{u^k + v^k} \stackrel{M}{\longrightarrow} 0$. \pend

\begin{lemma}
\label{LemDirClosure2} If $A \subseteq L^p_d$ and $X \in L^p_d$, then
\begin{equation}
\label{EqDirClosure} \cb{u \in M \colon X + u\One \in \cl_M A} =
    \cl\cb{u \in M \colon X + u\One \in A}.
\end{equation}
\end{lemma}

{\sc Proof.} Directly from definition \ref{DefRadCloSet}. \pend

\begin{remark}
If $A$ is closed in $L^p_d$ then it is obviously directionally closed in $M$ for each
possible $M$. This also holds for the $\sigma\of{L^\infty_d, L^1_d}$-topology on
$L^\infty_d$.
\end{remark}

\subsection{Translative functions on $L^p_d$}

The first result links the graph and the zero sublevel set of an $M$-translative
function.

\begin{proposition}
\label{PropGraph} A function $R \colon L^p_d \to \mathcal{P}\of{M}$ is translative in $M$
if and only if
\begin{equation}
\label{EqGrAR}
 \gr R = \cb{\of{X, u} \in L^p_d \times M \colon X + u\One \in A_R}.
\end{equation}
\end{proposition}

{\sc Proof.} Indeed, if $R$ is translative in $M$, then $\of{X, u} \in \gr R$ if and only
if $u \in \R\of{X}$ if and only if $0 \in R\of{X + u\One}$ by \eqref{EqTrans} if and only
if $X + u\One \in A_R$ by \eqref{EqAR}. On the other hand, if \eqref{EqGrAR} is true,
then $w \in R\of{X + u\One}$ for $u, w \in M$ if and only if $\of{X + u\One, w} \in \gr
R$ if and only if $X + \of{u+w}\One \in A_R$ by \eqref{EqGrAR} if and only if $\of{X , u
+ w} \in \gr R$ again \eqref{EqGrAR} if and only if $u+w \in \gr R$ which is
translativity in $M$. \pend

\medskip The following proposition has to be read as follows: The properties for $R$
and $A$ given in (a) through (m) produce two equivalencies, the first being that $R$ has
the property in question if and only if $A_R$ has the corresponding property, and the
second that $A$ has the property in question if and only if $R_A$ has the corresponding
property.

\begin{proposition}
\label{PropPrimalRepresProp} The following properties are in a one-to-one relationship
for an $M$-translative function $R \colon L^p_d \to \mathcal{P}\of{M}$ and a subset $A
\subseteq L^p_d$:
\\
(a) $R$ is $B$-monotone, and $A + B \subseteq A$;
\\
(b) $R\of{0} \neq \emptyset$, and $M\One \cap A \neq \emptyset$;
\\
(c) $R\of{0} \neq M$, and $M\One \cap \of{L^p_d \bs A} \neq \emptyset$;
\\
(d) $R$ maps into the set $\mathcal{P}_{K^M_I} = \cb{D \subseteq M \colon D = D +
K^M_I}$, and $A + K^M_I\One \subseteq A$;
\\
(e) $R$ is convex, and $A$ is convex;
\\
(f) $R$ is positively homogeneous, and $A$ is a cone;
\\
(g) $R$ is subadditive, and $A + A \subseteq A$;
\\
(h) $R$ is sublinear, and $A$ is a convex cone;
\\
(i) $R$ has closed images, and $A$ is directionally closed with respect to $M$;
\\
(k) $R$ has a closed graph, and $A$ is closed;
\\
(l) $R\of{X} \neq \emptyset$ for all $X \in L^p_d$, and $L^p_d = A + M\One$;
\\
(m) $R\of{X} \neq M$ for all $X \in L^p_d$, and $L^p_d = \of{L^p_d \bs A} + M\One$.

\end{proposition}

{\sc Proof.} We only give proofs for $A$ and $R_A$. The corresponding proofs for $R$ and
$A_R$ follow since $R = R_{A_R}$ and $A = A_{R_A}$ according to proposition
\ref{PropOneToOneB}.

(a) Assume $A + B \subseteq A$ and take $X^1, X^2 \in L^p_d$ such that $X^2 - X^1 \in B$.
Take $u \in R_A\of{X^1}$. Then $0 \in R_A\of{X^1 + u \One}$ by translativity and hence
$X^1 + u\One \in A_{R_A} = A$ (see proposition \ref{PropOneToOneB}). By assumption, $X^2
+ u\One = X^1 + u\One + X^2 - X^1 \in A$, hence $u \in R_A\of{X^2}$. This implies
$R_A\of{X^1} \subseteq R_A\of{X^2}$ as required. Vice versa, assuming the
$B$-monotonicity of $R_A$ and taking $X^1 \in A$, $X^2 \in B$ we obtain
\[
0 \in R_A\of{X^1} \subseteq R_A\of{X^1 + X^2}
\]
since $X^1 + X^2 - X^1 \in B$.

(b) and (c) are easily checked using the definition of $R_A$.

(d) Assume $A + K^M_I\One \subseteq A$, and take $u \in R_A\of{X}$, $u' \in K_I^M$. Then
$X + u\One \in A$. The assumption implies $X + \of{u + u'} \One \in A$, hence $u + u' \in
R_A\of{X}$ by (\ref{EqRA}). Assume $R_A$ maps into the set $\mathcal{P}_{K^M_I}$. Then $0
\in R_A\of{X + u\One} + u \subseteq R_A\of{X + u\One}$ whenever $X \in A$ and $u \in
K^M_I$ since $R_A$ is $M$-translative.

(e) -- (g) are easily checked.

(h) is a consequence of (f) and (g).

(i) Assume that $A$ is directionally closed in $M$. Take a sequence $\cb{u^k}_{k \in \N}
\subset R_A\of{X}$ with $\lim_{k \to \infty} u^k = u$. Then by (\ref{EqRA}) $X + u^k \One
=\of{ X + u\One} + \of{u^k - u}\One \in A$ for all $k \in \N$. Since $A$ is directionally
closed in $M$ this implies $X + u\One \in A$ which gives $u \in R_A\of{X}$. Conversely,
if $X \in L^p_d$, $u^k \stackrel{M}{\longrightarrow} 0$ and $X + u^k\One \in A$ then $0
\in R_A\of{X + u^k\One} = R_A\of{X} - u^k$ for all $k \in \N$. Since $R_A\of{X}$ is
closed $0 \in R_A\of{X}$, i.e. $X \in A = A_{R_A}$.

(k) Use proposition \ref{PropGraph}.

(l), (m) Immediate. \pend

\medskip The preceding proposition together with proposition \ref{PropOneToOneB} can be
used as a toolbox for one-to-one correspondences between classes of risk measures and
classes of acceptance sets. For example, closed convex risk measure which are finite at
zero are one-to-one with closed convex acceptance sets satisfying (A1a) and (A1b). We
also mention that proposition \ref{PropPrimalRepresProp} also includes the case $m=d=1$
which produces the well-known correspondence results for extended real-valued risk
measures and their acceptance sets in $L^p_1$ - even for $0 \leq p \leq \infty$.

\subsection{Convex set-valued functions and proof of the main theorem}

\label{SecAppConvex}

\noindent {\bf Fact 1.} $R \colon L^p_d \to \mathcal{P}\of{M}$ is convex if and only if
$\gr R$ is convex. The proof is straightforward.

\medskip\noindent {\bf Fact 2.} A $M$-translative function $R$ is convex if and only if $A_R$ is
convex if and only if all sets of the form
\[
\cb{X \in L^p_d \colon u \in R\of{X}}, \; u \in M,
\]
are convex. Proposition \ref{PropGraph} produces this fact.

\medskip\noindent {\bf Fact 3.} If $R$ is closed convex and $R\of{X^0} \neq M$ for $X^0 \in L^p_d$
with $R\of{X^0} \neq \emptyset$, then $R\of{X} \neq \emptyset$ for all $X \in L^p_d$. See
\cite{Hamel09}, proposition 5.

\medskip For the proof of theorem \ref{ThmDualRep}, we shall make use of the Fenchel conjugation
concept for set-valued functions introduced in \cite{Hamel09}, \cite{HamelHeyde10}. Let
$R \colon L^p_d \to \mathcal{P}\of{M}$ be a function. The Legendre--Fenchel conjugate of
$R$ is defined to be
\[
-R^*\of{Y, v} = \cl \bigcup_{X \in L^p_d} \of{R\of{X} + F^M_{\of{Y, v}}\sqb{-X}},
    \quad Y \in L^q_d, \; v \in M.
\]
The definition of the biconjugate depends on the order relation in the image space. For
$K_I$-compatible risk measures, the appropriate image space and the order relation will
be generated by the cone $K^M_I$. The following remark justifies the definition of the
biconjugate giving afterwards.

\begin{remark}
If $R \colon L^p_d \to \mathcal{P}\of{M}$ is a closed convex and $K_I$-compatible risk
measure, then $R\of{X} \in \mathbb{G}_M = \cb{D \subseteq \R^d \colon D = \cl\co\of{D +
K^M_I}}$ for all $X \in L^p_d$. In this case, the additional dual variable $v$ can be
restricted to $\of{K^M_I}^+$ (compare proposition \ref{PropExpectationRM}, (f)).
\end{remark}

The biconjugate of a function $R \colon L^p_d \to \mathbb{G}_M$ is defined by
\[
R^{**}\of{X} = \bigcap_{Y \in L^q_d, \, v \in \of{K^M_I}^+\bs\cb{0}}
    \of{-R^*\of{Y, v} + F^M_{\of{Y, v}}\sqb{X}}, \quad X \in L^p_d.
\]
Note that in these definitions the $F^M_{\of{Y, v}}$'s act as substitutes for real-valued
continuous linear functions. Note also that $v$ is an additional dual variable in
comparison to the scalar case which reflects the order in the image space as it will
become clear presently.

The following facts are basic results, see \cite[proposition 11]{Hamel09},
\cite[proposition 6.2]{HamelHeyde10}: We have $-R^* \colon L^q_d \times \of{K^M_I}^+ \to
\mathbb{G}_M$, and $R^{**}$ maps indeed into $\mathbb{G}_M$. The image $-R^*\of{Y, v}$ is
of the form $u + F^M_{\of{Y, v}}\sqb{0} = u + G\of{v}$ for some $u \in M$ or an element
of $\cb{M, \emptyset}$.

The Fenchel--Moreau theorem for set-valued functions holds true: Let $R \colon L^p_d \to
\mathbb{G}_M$, $1 \leq p \leq \infty$ be a function which is closed
($\sigma\of{L^\infty_d, L^1_d}$-closed in case $p=\infty$), convex, proper, then $R =
R^{**}$. Compare theorem 2 in \cite{Hamel09}.

It turns out that Fenchel conjugates for set-valued risk measures can be computed in
terms of set-valued support functions of their acceptance sets.

\begin{proposition}
\label{PropRiskMeasureConjugate} Let $R \colon L^p_d \to \mathbb{G}_M$ be a
market-compatible convex risk measure which is finite at zero and has the acceptance set
$A_R$. Take $Y \in L^q_d$ and $v \in \of{K^M_I}^+\bs\cb{0}$. Then
\begin{equation}
\label{EqRiskMeasureConjugate1} -R^*\of{Y, v} =
    \left\{\begin{array}{ccc}
    -S_{A_R}\of{Y, v} & : & Y \in -L^q_d\of{K^+_T}, \; v \in E\sqb{-Y} + M^\perp  \\
    M & : & \mbox{else}
    \end{array}\right.
\end{equation}
where
\[
-S_{A_R}\of{Y, v} =
    \cl \bigcup_{X \in A_R} F^M_{\of{Y, v}}\sqb{-X}.
\]
If $R$ is additionally positively homogeneous, then
\begin{equation}
\label{EqRiskMeasureConjugate2} -R^*\of{Y, v} =
    \left\{\begin{array}{ccc}
    G\of{v} \cap M & : & Y \in -A_R^+, \; v \in E\sqb{-Y} + M^\perp  \\
    M & : & \mbox{else}
    \end{array}\right.
\end{equation}
\end{proposition}

{\sc Proof.} Obviously,
\[
-R^*\of{Y, v} \supseteq \cl \bigcup_{X \in A_R} \of{R\of{X} + F^M_{\of{Y, v}}\sqb{-X}}
    \supseteq \cl \bigcup_{X \in A_R} F^M_{\of{Y, v}}\sqb{-X},
\]
hence $-R^*\of{Y, v} \supseteq -S_{A_R}\of{Y,v}$ for all $Y \in L^q_d$, $v \in \R^d$.

If $Y \not\in -L^q_d\of{K^+_T}$ then there is $X^0 \in L^p_d\of{K_T}$ such that
$E\sqb{Y^TX^0} > 0$. By (A1a) applied to $A_R$, there is $u^0 \in M$ such that $u^0\One
\in A_R$, hence $u^0\One + L^p_d\of{K_T} \subseteq A_R$. Using the definition of
$F^M_{\of{Y, v}}\sqb{-u^0\One -tX^0}$ we obtain
\[
-S_{A_R}\of{Y,v} \supseteq
    \cl \bigcup_{X \in u^0\One + L^p_d\of{K_T}} F^M_{\of{Y, v}}\sqb{-X}
    \supseteq \bigcup_{t > 0} F^M_{\of{Y, v}}\sqb{-u^0\One-tX^0} = M.
\]
Hence $-R^*\of{Y, v} \supseteq -S_{A_R}\of{Y,v} = M$ whenever $Y \not\in
-L^q_d\of{K^+_T}$.

Now, assume $E\sqb{Y} + v \not\in M^\perp$ and consider $w \in M$. Then
\begin{align*}
F^M_{\of{Y, v}}\sqb{- w\One} & =
    \cb{u \in M \colon 0 \leq v^Tu + E\sqb{Y}^Tw}
    \\ & = \cb{u - w \in M \colon 0 \leq v^T\of{u-w} + \of{E\sqb{Y} +v}^T w} + w
    \\ & = \cb{u \in M \colon 0 \leq v^Tu + \of{E\sqb{Y} + v}^T w} + w.
\end{align*}
Since $E\sqb{Y} + v \not\in M^\perp$ for each $u \in M$ we can find $w \in M$ such that
$0 \leq v^Tu + \of{E\sqb{Y} + v}^T w$. Therefore, it holds
\[
\bigcup_{w \in M} \of{F^M_{\of{Y, v}}\sqb{-w\One} -  w} = M
\]
and a fortiori, since $R\of{0} \neq \emptyset$ by (R1a),
\begin{align*}
-R^*\of{Y, v} & =
    \cl \bigcup_{X \in L^p_d, w \in M} \of{R\of{X+w\One} + F^M_{\of{Y, v}}\sqb{-X-w\One}}
    \\
    & \supseteq \cl \bigcup_{w \in M} \of{R\of{0} - w + F^M_{\of{Y, v}}\sqb{-w\One}}
    = M.
\end{align*}

It remains to show that $-R^*\of{Y, v} \subseteq -S_{A_R}\of{Y,v}$ for $Y \in
-L^q_d\of{K^+_T}$, $E\sqb{Y} + v \in M^\perp$. Indeed, taking $u \in R\of{X}$ with $X \in
\dom R$ we obtain $X + u\One \in A_R$ and with Proposition \ref{PropExpectationRM}
(applied with $Y$ replaced by $-Y$)
\[
-S_{A_R}\of{Y,v} \supseteq F^M_{\of{Y, v}}\sqb{-X-u\One} = F^M_{\of{Y, v}}\sqb{-X} + u.
\]
Therefore, $R\of{X} + F^M_{\of{Y, v}}\sqb{-X} \subseteq -S_{A_R}\of{Y,v}$ for all $X \in
L^p_d$, and hence $-R^*\of{Y, v} \subseteq -S_{A_R}\of{Y,v}$.

Now, let $R$ be additionally positively homogeneous. If $Y \not\in -A^+_R$ then there is
$X^0 \in A_R$ such that $E\sqb{Y^TX^0} > 0$ which yields $\bigcup_{X \in A_R} F^M_{\of{Y,
v}}\sqb{-X} \supseteq \bigcup_{t > 0} F^M_{\of{Y, v}}\sqb{-tX^0} = M$, and a fortiori
$-R^*\of{Y, v} = M$ by \eqref{EqRiskMeasureConjugate1}. If $Y \in -A^+_R$ then
$E\sqb{Y^TX} \leq 0$ for all $X \in A_R$, hence $F^M_{\of{Y, v}}\sqb{-X} \subseteq
F^M_{\of{Y, v}}\sqb{0}$ for all $X \in A_R$ and therefore, since $0 \in A_R$
\[
F^M_{\of{Y, v}}\sqb{0} \subseteq \bigcup_{X \in A_R} F^M_{\of{Y, v}}\sqb{-X} \subseteq
F^M_{\of{Y, v}}\sqb{0}.
\]
This completes the proof of the proposition. \pend

\medskip The reader may note that formula \eqref{EqRiskMeasureConjugate2} gives the
conjugate of a sublinear risk measure in terms of a set-valued indicator function if one
agrees to interpret $G\of{v} \cap M$ as "zero" and $M$ as "$-\infty$". This
interpretation, a perfect analogy to the scalar case, can be made more precise. See
\cite{Hamel09} for more details.

\begin{remark}
\label{RemPreAcceptSupport} Let $\widehat{A} \subseteq L^p_d$ be a convex acceptance set.
Then, the set $A = \cl\of{\widehat{A} + K^M_I\One}$ is closed convex and
\[
\forall Y \in L^q_d, \; \forall v \in \of{K^M_I}^+\bs\cb{0} \colon
    -S_{A}\of{Y, v} = -S_{\widehat{A}}\of{Y, v}.
\]
Indeed, while "$\supseteq$" is trivial the converse inclusion follows from the fact that
\[
F^M_{\of{Y, v}}\sqb{-X} + K^M_I \subseteq F^M_{\of{Y, v}}\sqb{-X}
\]
since $v \in \of{K^M_I}^+$ and a continuity argument. This observation can also be seen
as an analogy to a well-known scalar fact: The extended real-valued support function of a
set coincides with the support function to the closure of the convex hull of this set.
\end{remark}

{\sc Proof of theorem \ref{ThmDualRep}.} Assume that $R$ is a closed convex risk measure
which is finite at zero. From the Fenchel--Moreau theorem for set-valued functions (see
\cite{Hamel09}, theorem 2) and proposition \ref{PropRiskMeasureConjugate} we obtain
\begin{equation}
\label{EqRMDual1} R\of{X} =
    \bigcap_{Y \in -L^q_d\of{K^+_T}, \, v \in \of{E\sqb{-Y} + M^\perp}\cap \of{K^M_I}^+\bs\cb{0}}
    \sqb{-S_{A_R}\of{Y, v} + F^M_{\of{Y, v}}\sqb{X}}.
\end{equation}
By lemma \ref{LemDualTransform} we have $Y \in -L^q_d\of{K^+_T}$ and $v \in \of{E\sqb{-Y}
+ M^\perp}\cap \of{K^M_I}^+\bs\cb{0}$ if and only if we can find $\of{Q, w} \in
\mathcal{W}^q$ such that
\begin{equation}
\label{EqRMDual2}
 \forall X \in L^p_d \colon
    \widetilde{F}^M_{\of{Q, w}} \sqb{X} = F^M_{\of{Y, v}}\sqb{-X}
\end{equation}
since $F^M_{\of{-Y, v}}\sqb{X} = F^M_{\of{Y, v}}\sqb{-X}$. For pairs $\of{Y, v}$,
$\of{Q,w}$ satisfying \eqref{EqRMDual2} we obtain
\[
-S_{A_R}\of{Y, v} = \cl \bigcup_{X \in A_R} F^M_{\of{Y, v}}\sqb{-X} =
    \cl \bigcup_{X \in A_R} F^M_{\of{-Y, v}}\sqb{X} =
    \cl\bigcup_{X \in A_R} \widetilde{F}^M_{\of{Q, w}}\sqb{X}.
\]
Defining a function $-\widetilde{S}_{A_R} \colon \mathcal{W}^q \to \mathbb{G}_M$ by
\[
-\widetilde{S}_{A_R}\of{Q, w} = \cl\bigcup_{X \in A_R} \widetilde{F}^M_{\of{Q, w}}\sqb{X}
\]
we get
\[
-S_{A_R}\of{Y, v} + F^M_{\of{Y, v}}\sqb{X} =
    -\widetilde{S}_{A_R}\of{Q, w} + \widetilde{F}^M_{\of{Q, w}}\sqb{-X}
\]
whenever $\of{Y, v}$, $\of{Q,w}$ satisfy \eqref{EqRMDual2}. This gives
\eqref{ThmDualRep1} with $-\alpha_R = -\alpha_{R, \min} =  -\widetilde{S}_{A_R}$ because
$\widetilde{F}^M_{\of{Q, w}}\sqb{-X} = \of{E^Q\sqb{-X} + G\of{w}}\cap M$ (see lemma
\ref{LemDualTransform}).

The function $-\alpha_{R, \min} = -\widetilde{S}_{A_R}$ indeed has all the properties of
a penalty function. The very definition of the $\widetilde{F}^M_{\of{Q, w}}$'s ensures
that $-\widetilde{S}_{A_R}$ maps into $\mathbb{G}_M$. The properties in (P0) follow from
$R\of{0} \neq \emptyset$ and $R\of{0} \neq M$, respectively. (P1) is a consequence of
\[
-\widetilde{S}_{A_R}\of{Q, w} + G\of{w} =
    \cl\bigcup_{X \in A_R} \widetilde{F}^M_{\of{Q, w}}\sqb{X} + G\of{w}
    \subseteq \cl\bigcup_{X \in A_R} \widetilde{F}^M_{\of{Q, w}}\sqb{X}.
\]

Finally, if \eqref{ThmDualRep1} is satisfied for a penalty function $-\alpha_R$ then
\[
\forall X \in L^p_d, \; \forall \of{Q, w} \in \mathcal{W}^q \colon
    R\of{X} \subseteq
    -\alpha_R\of{Q, w} + \widetilde{F}^M_{\of{Q, w}}\sqb{X}.
\]
Adding $\widetilde{F}^M_{\of{Q, w}}\sqb{-X}$ on both sides of the inclusion and taking
the union over $X \in L^p_d$ on the left hand side we obtain in view of the properties of
a penalty function
\[
\cl\bigcup_{X \in L^p_d} \of{R\of{X} + \widetilde{F}^M_{\of{Q, w}}\sqb{-X}}
    \subseteq -\alpha_R\of{Q, w}.
\]
The left hand side of this inclusion is nothing else then the conjugate of $R$ expressed
in terms of $\of{Q, w}$ instead of $\of{Y, v}$. Via lemma \ref{LemDualTransform} we get
with the help of proposition \ref{PropRiskMeasureConjugate} $-\alpha_{R, min}\of{Q, w}
\subseteq -\alpha_R\of{Q, w}$ for all $\of{Q, w} \in \mathcal{W}^q$.

Vice versa, one may check directly that $R$ generated via \eqref{ThmDualRep1} by a given
penalty function $-\alpha_R$ is a closed convex risk measure.

The coherent case follows with the help of \eqref{EqRiskMeasureConjugate2}. \pend

\end{document}